\documentclass[preprint,10pt]{elsarticle}

\usepackage{graphicx}

\usepackage{natbib}
\usepackage{colortbl}
\usepackage{amssymb}
\usepackage{moreverb} 
 \usepackage{float}   
\usepackage{listings}
\usepackage[linesnumbered,ruled]{algorithm2e}
\SetCommentSty{mycommfont}
\usepackage{nomencl}

\usepackage[english]{babel}

\usepackage{nomencl}  
\usepackage{leftidx} 
 \usepackage{subfigure}
 \usepackage{amsmath, amsthm, amssymb}

\usepackage[left=1.5cm,top=2cm,right=1.5cm,bottom=2.5cm]{geometry} %
\usepackage{color}
\usepackage{listings}
\usepackage{mathrsfs}

\newtheorem{remark}{Remark}[section]

\usepackage{multirow}
\usepackage{color,colordvi,graphicx,xcolor}
\DeclareMathOperator*{\argmax}{arg\,max} 

\definecolor{mygreen}{rgb}{0,0.6,0}
\definecolor{mygray}{rgb}{0.5,0.5,0.5}
\definecolor{mymauve}{rgb}{0.58,0,0.82}
\usepackage{hyperref}
\makenomenclature

\usepackage{lipsum}
\usepackage{tikz}
\usepackage{bbm}

\usepackage{import}
\usepackage{xifthen}
\usepackage{pdfpages}
\usepackage{transparent}

\newcommand{\norm}[1]{\left\lVert#1\right\rVert} 
\usepackage{optidef} 
\usepackage{tcolorbox}

\newtcolorbox{mybox2}[1][]{
    colback=white,       
    #1                   
}

\usepackage{newfloat}
\DeclareFloatingEnvironment[
    fileext=lob,
    listname={List of Boxes},
    name=Box,
    placement=tbhp,
    within=section, 
]{mybox}

\usepackage{xcolor}

\newcommand{\revone}[1]{\textcolor{black}{#1}}   
\newcommand{\revtwo}[1]{\textcolor{black}{#1}}    
\newcommand{\own}[1]{\textcolor{black}{#1}}    

\begin{document}

\begin{frontmatter}

\title{A  subspace-adaptive weights   cubature method     with application to   the local hyperreduction of parameterized finite element models}

\author[upc,cimne]{J.R. Bravo\corref{cor1}}
\ead{jose.raul.bravo@upc.edu}
\author[upc-terrasa,cimne]{J.A. Hern{\'a}ndez}
\author[upc,cimne]{S. Ares de Parga}
\author[upc,cimne]{R. Rossi}

%
\cortext[cor1]{Corresponding author}

\address[upc]{Universitat Polit\`{e}cnica de Catalunya, Department of Civil and Environmental Engineering (DECA), Barcelona, Spain}
\address[cimne]{Centre Internacional de M\`{e}todes Num\`{e}rics en Enginyeria (CIMNE), Barcelona, Spain}
\address[upc-terrasa]{Universitat Polit\`{e}cnica de Catalunya,  E.S. d'Enginyeries Industrial, Aeroespacial i Audiovisual de Terrassa (ESEIAAT), Terrassa, Spain}%

\begin{abstract}

This  paper is     concerned with quadrature/cubature   rules able to deal   with  multiple subspaces of   functions, in such a way    that   the    integration points are common  for all the subspaces, yet  the (nonnegative)  weights are tailored to each specific   subspace.  These  subspace-adaptive weights  cubature rules can be used to accelerate  computational mechanics applications   requiring    efficiently  evaluating spatial integrals  whose integrand function   dynamically switches      between multiple pre-computed subspaces. One of such applications is   local hyperreduced-order modeling (HROM), in which the solution  manifold is approximately represented  as a collection of   basis matrices, each basis matrix corresponding to a different region in    parameter space.  The proposed optimization framework is discrete in terms of the location of the integration points, in the sense that such points are selected among the Gauss points of a given finite element mesh, and the target subspaces of functions are represented by orthogonal basis matrices constructed from the values of the functions at such Gauss points, using the Singular Value Decomposition (SVD). This discrete framework allows us to treat also problems in which the integrals are approximated as a weighted sum of the  contribution of each finite element, as in the     Energy-Conserving Sampling and Weighting (ECSW) method of C. \own{Farhat} and co-workers.  Two distinct solution strategies are examined. The first one is a  greedy strategy based on an enhanced version of the Empirical Cubature Method (ECM) developed by the authors elsewhere (we call it the Subspace-Adaptive Weights ECM, SAW-ECM for short), while the  second method is based on a convexification of the cubature problem so that it can be addressed by  linear programming algorithms. We show in a toy problem  involving integration of polynomial functions that the SAW-ECM clearly outperforms the other method both in terms of computational cost and optimality. On the other hand, we illustrate the performance of the SAW-ECM in  the construction of a local HROMs in a highly nonlinear equilibrium   problem (large strains regime).  We demonstrate that, provided that the subspace-transition errors are negligible,  the error associated to hyperreduction using adaptive weights can be controlled by the truncation tolerances of the SVDs used for determining the basis matrices.    We also show that the number of integration points decreases notably as the number of subspaces increases, and that,  in the limiting case of using as many subspaces as snapshots, the SAW-ECM delivers   rules with a number of integration  points only dependent on the intrinsic dimensionality of the solution manifold and the degree of overlapping required to avoid subspace-transition errors.
The Python source codes of   the proposed SAW-ECM  are openly accessible in the public
repository \url{https://github.com/Rbravo555/localECM}.

\end{abstract}
\end{frontmatter}

\newcommand{\E}{\boldsymbol{E}}
 \renewcommand{\P}{\boldsymbol{P}}

\renewcommand{\d}{\boldsymbol{d}}
\newcommand{\betaB}{\boldsymbol{\beta}}

\newcommand{\x}{\boldsymbol{x}}
\newcommand{\w}{\boldsymbol{w}}
\newcommand{\ngausT}{M}
\newcommand{\pNMOD}{m}
\newcommand{\xG}[1]{\bar{\x}_{#1}}
\newcommand{\xBAR}{\bar{\x}}

\newcommand{\xDECM}{\xBAR}

 \newcommand{\nSUBS}{k}
 
\renewcommand{\b}{\boldsymbol{b}}
\renewcommand{\r}{\boldsymbol{r}}
\newcommand{\s}{\boldsymbol{s}}
\newcommand{\G}{\boldsymbol{G}}
\renewcommand{\H}{\boldsymbol{H}}

\renewcommand{\u}{\boldsymbol{u}}
\newcommand{\Ui}[1]{\boldsymbol{U}^{(#1)}}
\newcommand{\dI}[1]{\boldsymbol{d}^{(#1)}}
\newcommand{\bI}[1]{\boldsymbol{b}^{(#1)}}

\newcommand{\zetaI}[1]{\boldsymbol{\zeta}^{(#1)}}
\newcommand{\zetaB}{\boldsymbol{\zeta}}

 \newcommand{\ident}{\boldsymbol{I}} 

  \newcommand{\intG}[3]{\displaystyle \int_{#1}  \,#3  \,\, d#2}  

 \newcommand{\paramSPC}{\mathcal{D}}    
 \newcommand{\f}{\boldsymbol{f}}
  \newcommand{\q}{\boldsymbol{q}}
  \newcommand{\V}{\boldsymbol{V}}
    \newcommand{\y}{\boldsymbol{y}}
    \newcommand{\mP}{m}
    \newcommand{\omegaB}{\boldsymbol{\omega}}
\renewcommand{\refeq}[1]{Eq.(\ref{#1})} 

    \newcommand{\A}{\boldsymbol{A}}
    \newcommand{\W}{\boldsymbol{W}}
    \newcommand{\U}{\boldsymbol{U}}
    \newcommand{\z}{\boldsymbol{z}}
 \newcommand{\zero}{\boldsymbol{{0}}}

  \newcommand{\F}{\boldsymbol{F}}
    \newcommand{\Eind}{\mathcal{E}}

  \newcommand{\K}{\boldsymbol{K}}
    \newcommand{\B}{\boldsymbol{B}}
   \newcommand{\ORDER}[1]{\mathcal{O}(#1)}

  \newcommand{\R}{\boldsymbol{R}}
  \newcommand{\PhiB}{\boldsymbol{\Phi}}
  \newcommand{\dTILDE}{\boldsymbol{\tilde{d}}}

 \newcommand{\Rn}[1]{\mathbb{R}^{#1}}  
 \newcommand{\inpP}{\boldsymbol{\mu}}
  \newcommand{\refpar}[1]{(\ref{#1})} 
 \newcommand{\Par}[1]{\left( #1 \right)}
 \newcommand{\fINT}{\boldsymbol{f}_{\!\textbf{int}}}
  \newcommand{\kINT}{\boldsymbol{k}_{\!\textbf{int}}}

 \newcommand{\Fint}{\boldsymbol{F}_{\!\textbf{int}}}
  \newcommand{\RRn}[2]{\mathbb{R}^{#1 \times #2}}  
  \newcommand{\NNn}[2]{\mathbb{N}^{#1 \times #2}}  
  
    \newcommand{\jahoHIDE}[1]{ }  

  \newcommand{\spanb}[1]{ {span}( #1)   }

\section{Introduction}
\label{section:introduction}

\subsection{Cubature rules for multiple   subspaces}
\label{sec:cubat.}

The classical  problem   known in  numerical analysis as  quadrature (for one-dimensional  domains) or cubature (for multidimensional   domains) of functions  pertaining to a given subspace of integrable functions on a domain $\Omega \subset  \Rn{d}$    consists in finding the smallest set of   points $\{ \x_1,\x_2 \ldots \x_m \}$   and   associated (positive)  scalar weights $\{w_1,w_2 \ldots w_m\}$ such that
\begin{equation}
\label{eq:001}
   \intG{\Omega}{\Omega}{\f(\x )} =  \sum_{g=1}^m \f(\x_g ) w_g
\end{equation}
is satisfied for any $\f$ pertaining to that subspace.  For instance, the optimal quatrature rule for the subspace of  univariate polynomials  up to degree $5$  in $\Omega = [-1,1]$  is the   Gaussian rule with weights $w_1=w_3 =5/9$, $w_2 = 8/9$ and positions $\x_1 = -\x_3 = \sqrt{3/5}$ and $x_2 = 0$  \cite{golub2009matrices}.

The present paper is     concerned with quadrature/cubature problems, but of a   distinct flavor from the classical one presented above:  instead      of    considering a single subspace of  integrable functions,  our goal is to derive    rules able to deal   with  multiple subspaces of integrable functions, in such a way    that   the  locations of the   integration points are common  for all the subspaces, yet    each subspace has a distinct set of    weights associated to such positions.  This generalized quadrature/cubature problem    may be   posed as follows: given    $P$ different   subspaces of integrable functions on $\Omega$,  find the smallest set of points $\{\x_1,\x_2 \ldots \x_m\}$   and   the associated nonnegative  weights for each subspace  $\{w_1^1, w_2^1, \ldots w_m^1\}$, $\{w_1^2, w_2^2, \ldots w_m^2\}$ $\ldots$ $\{w_1^P,w_2^P \ldots  w_m^P\}$    such that
\begin{equation}
\label{eq:002}
   \intG{\Omega}{\Omega}{\f^j(\x)} =  \sum_{g=1}^m \f^j(\x_g) w^j_g
\end{equation}
  holds for  any $\f^j$   ($j=1,2 \ldots P$);  here $\f^j$ denotes an arbitrary function of the $j$-th subspace.

Notice that a feasible solution of the above problem
 can be obtained by solving the standard cubature problem \refpar{eq:001} for the sum of all the subspaces ---we will refer hereafter to this sum of subspaces as the global subspace.   However,   this  feasible solution is far from being optimal, because the number of required points   scales with the dimension of the global subspace.    To illustrate this point, let us return to  the  example put forward above for the integration of monomials up to degree $5$ on $\Omega = [-1,1]$.  We know that  the  aforementioned 3-point Gaussian rule   is able to exactly integrate any linear combination of  such monomials. But now consider the (toy)   problem of deriving an integration rule for each monomial separately ($f^j =\alpha {x^{j}}$, $j=0,1,2 \ldots 5$, $\alpha \in \Rn{}$).    Since $ \int_{-1}^{1} \alpha {x^{j}}\,\, dx  =  \alpha (1- (-1)^{j+1})/({1+j})  $, it follows that   any one-point quadrature rule with  position $x\in ]0,1]$ (the same for all the six monomials) and   weights $w^j = 2/(x^{j+1}(1+j))$, for $j=0,2,4$, and $w^j =0$  for $j=1,3,5$,   is an optimal solution of problem \refpar{eq:002}.  Thus,  by tailoring    the weights  for each of the 6 one-dimensional subspaces (the monomials),   we arrive at   a quadrature rule with just  one point,   two points \own{less than} if we consider a rule for the global subspace. It is easy to see that  this conclusion can be  extended to any polynomial degree, that is,  the number of points of the subspace-adaptive weights  rule will be always equal to one, whereas the number of integration points when using the global Gaussian rule will scale linearly with the number of monomials (i.e., with the dimension of the  global subspace).

 \subsection{Applications in hyperreduced-order modeling}
 \label{sec:j,lllodddd}

 This apparent advantage  of   subspace-adaptive weights cubature rules (number of integration points independent of the dimension of the global subspace)  can be leveraged in  computational mechanics applications   in which   it is necessary to efficiently  evaluate integrals  whose integrand function   dynamically switches      between different pre-computed subspaces.  One of such applications ---actually the one that motivates the present work---  is the construction of (projection-based) \emph{local} \emph{hyperreduced-order models (HROMs)}  for parameterized finite element   problems.  The qualifier \emph{local} here is used in the sense given in the   work of Amsallem et al.\cite{amsallem2012nonlinear}, according to which a projection-based, reduced-order model is local when the solution  manifold is approximately represented  as the sum of the spans of multiple basis matrices, each basis matrix corresponding to a different region in    parameter space.  On the other hand, the term hyperreduction  indicates that, aside from the inherent reduction in number of equations  achieved by the projection  onto low-dimensional subspaces, there is an additional step of dimensionality reduction; this  steps  consists in computing the nonlinear terms of the projected equations using only a relatively small  subset of finite elements/Gauss points of the underlying finite element mesh \cite{hernandez2017dimensional, farhat2014dimensional}.   It is precisely in this hyperreduction   step where the usefulness of a subspace-adaptive weights cubature rule of the form \refeq{eq:002} becomes evident.

 Indeed,    consider, for the sake of concreteness, the   case  of  an HROM for a nonlinear structural  finite element problem (using Galerkin projection) in which the only   term requiring hyperreduction is the projected nodal internal forces.
 Let us denote by $\PhiB^1$, $\PhiB^2$ \ldots $\PhiB^P$  the   reduced basis matrices ($\PhiB^i \in \RRn{N}{n^i}$, $n^i << N$), obtained in the offline training stage from a set of representive solutions  by means of, say, a combination of  clustering algorithms such as k-means \cite{brunton2019data} with the truncated Singular Value Decomposition (SVD). Suppose that a given instant   of the online analysis the nodal displacements $\d \in \Rn{N}$ lies in the subspace corresponding to the $i$-th basis matrix, i.e.,   $\d = \PhiB^{i} \q^{i}$, $\q^i \in \Rn{n^i}$ being the reduced coordinates.  Computing the reduced internal forces at this instant requires evaluating the following  integral over the spatial domain $\Omega$
 \begin{equation}
 \label{eq:003}
  \F^{i} =  \intG{\Omega}{\Omega}{{\PhiB^{i}}^T \fINT(\PhiB^i \q^i)}
 \end{equation}
 where $\fINT$ denotes the internal force per unit volume   in its globally supported format.    The key observation here is that the subspace in which the  integrand function $\f^i = {\PhiB^{i}}^T \fINT$ resides can be pre-computed in the offline stage for each $i=1,2 \ldots P$ ---in fact, this is the same observation upon which standard global HROMs are based \cite{hernandez2017dimensional,farhat2014dimensional}, the only difference being that now we have to pre-compute $P$ different subspaces.  It follows then  that, as asserted, the  evaluation of the reduced internal forces \refeq{eq:003}  in a local HROM  falls   within the category of cubature problems in which the integrand function changes between pre-computed subspaces, and consequently,      cubature rules of the form \refpar{eq:002} can be legitimately applied.

 Other hyperreduction  problems that can benefit from using cubature rules with subspace-adaptive weights are those in which the residual   is formed by the sum of more than one nonlinear term. This is the case of   Least-Squares Petrov-Galerkin HROMs, in which the left basis matrix onto which the FE residual is projected changes at each iteration of the   analysis,  and therefore, all the  terms of such a residual   require hyperreduction \cite{grimberg2021mesh}.   Rather than treating all the contributions with the same cubature rule, as done e.g. in  Refs. \cite{grimberg2020stability,ares2023hyper,grimberg2021mesh,tezaur2022robust},  the cubature rules developed herein   would allow the separate  hyperreduction of   each nonlinear term, using different weights yet   the same set of points/elements. By the same logic exposed for the toy problem with polynomials, this set will be necessarily smaller than when approximating all the contributions with the same weights.

 \subsection{Methodology:  \own{s}ubspace-\own{a}daptive \own{w}eights   \own{c}ubature  }
 \label{sec:LCO}

 The methodology employed here for deriving subspace-adaptive cubature rules draws on previous works by the authors on the approach termed the \emph{Empirical Cubature Method} (ECM). The  ECM   was originally proposed  in     Ref. \cite{hernandez2017dimensional}   for global  Galerkin   HROMs (i.e., cubature problems of the form \ref{eq:001});  improved algorithmically later on in Ref.\cite{hernandez2020multiscale} by introducing rank-one updates in the iterative procedure; and extended to global HROMs based on least-squares Petrov-Galerkin projections  in Ref.\cite{ares2023hyper}, and to continuous cubature  in Ref. \cite{hernandez2024cecm}.  The ECM  in the first three works treats the cubature problem as a constrained \emph{best subset selection} problem, which means that the reduced set of integration points is assumed to be a subset  of the Gauss points of the underlying FE mesh.  Thus, the particular form of the functions to be integrated is of no concern in this discrete ECM, the only required information being their values at the $M$ Gauss points of the underlying mesh.  More specifically,  the target subspace of integrable functions is represented  by an orthogonal  basis matrix   $\U \in \RRn{M}{m}$,  computed from the snapshots of the integrand  using the truncated SVD (the truncation tolerance controls thus the integration error).   The ECM returns a set of $m$ indices $\Eind \subset \{1,2 \ldots M \}$ (i.e., as many indices as basis functions) and positive weights $\w\in \Rn{m}$ so that $\U_{\Eind}^T \w = \U^T \W$;   here $\U_{\Eind} = \U(\Eind,:)$ denotes the (square) block matrix of $\U$ formed by the rows with indices  $\Eind$,  $\W$ is the vector of (strictly positive ) FE integration weights   and, thus, $\U^T \W$ represents the   integrals  of the $m$ basis functions. Note that this problem is ill-posed when $\U^T \W \approx \zero$;  to avoid this, in the ECM $\U$ is augmented with an extra column so that its span  includes the vector space of constant functions while preserving orthogonality  \cite{hernandez2024cecm}.

 We will also adopt     this  discrete framework for solving the   cubature problem with subspace-adaptive weights. Accordingly,     the target subspaces   appearing in   problem \ref{eq:002} will be   characterized  by $P$ distinct orthogonal basis matrices $\U^1$, $\U^2$, \ldots $\U^{P}$  ($\U^i \in \RRn{M}{m^i}$), and   the     cubature problem  will be rephrased  as that       of finding the smallest set of  indices $\Eind$ and   nonnegative weights vectors $\w^1$, $\w^2$ \ldots $\w^P$ such that ${\U^i_{\!\Eind}}^T \w^i = {{\U^i}}^T \W$ holds for all $i =1,2 \ldots P$. It should be noted   that this discrete formulation can also   accommodate the  case in which the  integral is approximated  as a weighted sum of  the contributions of a subset of the $E$ finite elements of the mesh, i.e.:
\begin{equation}
\label{eq:002aa}
     \sum_{e=1}^E  \intG{\Omega^e}{\Omega}{\f^j(\x)} =    \sum_{g\in \Eind} w_g^j \Par{\intG{\Omega^g}{\Omega}{\f^j(\x)}}, \hspace{1cm} j = 1,2 \ldots P
\end{equation}
 (one just needs to replace the vector of Gauss weights $\W$ by an all-ones vector, and construct the  basis matrices $\U^i$  from the information of the element contributions). Thus, the  proposed methodology may be viewed as well as a  \emph{mesh sampling} procedure, like the Energy-Conserving   Sampling and Weighting (ECSW) procedure of Farthat and co-workers in Ref. \cite{farhat2014dimensional}, but now able to cope with multiple subspaces ---and  since the weights in our formulation are forced to be nonnegative, our   scheme will be also ``energy-conserving'', in the sense defined in Ref. \cite{farhat2015structure}.

 Two distinct methods for solving the best subset selection  problem described above will be examined.  The first method, that we will call the \emph{subspace-adaptive weights ECM}  (abbreviated SAW-ECM),  is a greedy algorithm   that involves sequentially applying  an enhanced version of the  standard ECM described in the foregoing  to each basis matrix.      At  each step, this enhanced ECM attempts to solve the local cubature problem of a given basis matrix $\U^i$   using as candidate  indices those obtained in previous steps. If it does not succeed, the algorithm    proceeds by expanding the  candidate set with the remaining indices, and repeats    this operation   until     all  local problems are solved.

The second method may be regarded as the generalization of the  linear program-based empirical quadrature    proposed by Patera and co-workers in \cite{patera2017lp,yano2019lp}. It  capitalizes  on the fact  that, as it occurs with the standard  cubature problem (see discussion in the introductory section of Ref. \cite{hernandez2024cecm}),   this multiple weights problem  can be  cast as a nonconvex sparsification problem in which the objective function is defined in terms of the $\ell_0$ pseudo-norm \cite{elad2010sparse}; this nonconvex problem  can be in turn convexified by  replacing  this pseudo-norm by the $\ell_1$ norm,   resulting in an optimization problem which is linear in both the objective functions and the constraints and, therefore, amenable to be solved by standard linear progamming.

We will show in a   quadrature problem of polynomial functions \own{(S}ection \ref{sec:examp1}) that    the  {SAW-ECM} largely outperforms the other  approach based on linear programming, both in terms of computational efficiency and optimality with respect to the number of points. For this reason,  the assessment of the methodology in the context of local HROMs (Section \ref{sec:localH}) will be carried out solely using this second approach.  The goal of this assessment will be twofold. On the one hand, we will investigate whether, as it occurs with the standard ECM \cite{hernandez2017dimensional}, the error introduced by the SAW-ECM can be  controlled by the truncation tolerances of the SVDs used to construct the local basis matrices $\U^1, \U^2 \ldots \U^P$. On the other hand, we will  analyze how   the number of integration points  decreases as  the partition of the solution space becomes finer. Of special interest  will be to examine what occurs in  the limiting case in which the number of subspaces  is (almost)  equal to the number of snapshots used to construct the basis matrices, for
it represents   the scenario in which the number of integration points will be reduced the most in comparison with the global approach of using the same weights for all  subspaces.

 \subsection{Originality of this work}
\label{sec:origin}
\begin{figure}[!ht]
    \centering
    \includegraphics[width=.7\linewidth]{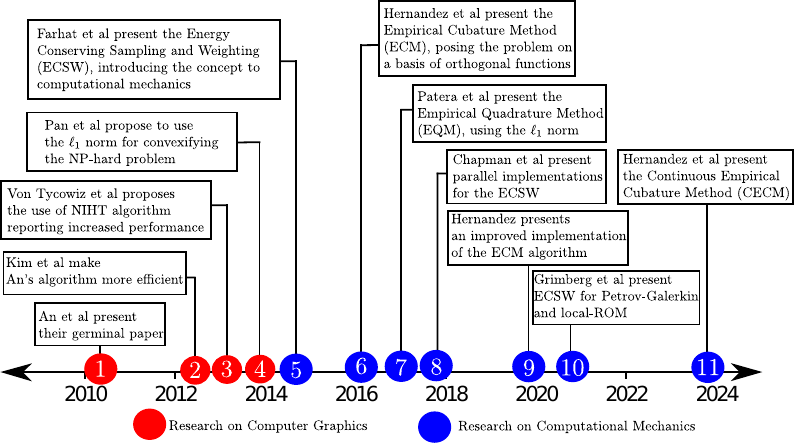}
    \caption{ {Summary of the state-of-the-art review on    cubature rules/mesh sampling procedures presented in Ref. \cite{hernandez2024cecm}. The references on the timeline are \textbf{1)} An et al \cite{an2009optimizing}, \textbf{2)} Kim et al \cite{kim2013subspace}, \textbf{3)} Von Tycowiz et al \cite{von2013efficient},  \textbf{4)} Pan et al \cite{pan2015subspace},   \textbf{5)} Farhat et al \cite{farhat2014dimensional}, \textbf{6)} Hernandez et al \cite{hernandez2017dimensional}, \textbf{7)} Patera et al \cite{patera2017lp}, \textbf{8)} Chapman et al \cite{chapman2016accelerated}, \textbf{9)} Hernandez \cite{hernandez2020multiscale}, \textbf{10)} Grimberg et al \cite{grimberg2021mesh}, \textbf{11)} Hernandez et al \cite{hernandez2024cecm}}}
    \label{fig: time line cubature}
\end{figure}

The authors carried out in a  recent paper  \cite{hernandez2024cecm} a comprehensive review of cubature-based hyperreduction methods, from its inception in the computer graphic community with  the seminal paper by An et al. \cite{an2009optimizing}, to its appearance \own{(w}ith the ECSW of Farhat and co-workers \cite{farhat2014dimensional}) and subsequent expansion in computational engineering circles.  A  time-line summary of this review is shown  in Figure \ref{fig: time line cubature}.  The pioneering paper by Amsallem et al.  in 2012   \cite{amsallem2012nonlinear}, which is     the first work combining local subspaces   and hyperreduction,  and later refinements in this research strand \cite{amsallem2015fast}, are not included in this time-line graph because Amsallem's works do not employ for hyperreduction a cubature-based method, but rather  an ``approximate-then-project'' approach   \own{(}cubature methods are project-then-approximate methods, see classification given in   \cite{farhat2020computational}).

The first paper in the survey summarized in Figure \ref{fig: time line cubature}      that explicitly raises the question of how to efficiently deal with cubature-based methods (more specifically, the  ECSW) in which   there are  several terms requiring hyperreduction (as in the alluded to earlier Least-Squares Petrov-Galerkin HROMs) or/and several local subspaces (as in local HROMs) is the work by Grimberg et al. in 2021 \cite{grimberg2021mesh}.  However, Grimberg et al. only consider two possible scenarios: a global cubature/mesh sampling method (they call it ``grouped hyperreduction''), or ``individual'' hyperreduction ---and by individual they mean each term/subpsace having its own cubature rule, with different weights and different points. Thus, Grimberg et al.  \cite{grimberg2021mesh} do not contemplate the possibility of using different weights but shared  points.   A broader review of the literature on  disciplines dealing with the efficient numerical evaluation of integrals (such as the field of Generalized Gaussian rules, see e.g. \cite{kovvali2022theory} and references therein) has not revealed  any approach dealing with the notion of integration rules for multiple subspaces of functions  using subspace-adaptive weights and common points either. Thus,
to the best of the authors' knowledge,  not only the idea of using  subspace-adaptive weights cubature rules in   hyperreduced-order models,  but the formulation and solution of the problem itself    are  original contributions of the present work.

\revone{
\begin{remark}
The literature review presented in the foregoing is   incomplete insofar as hyperreduction is concerned, for it only takes into consideration hyperreduced schemes based on cubature/energy-conserving mesh sampling procedures in which the solution manifold is approximated in a piecewise linear fashion ---and thus ignores    the numerous approaches published in recent years that use more advanced representations of the solution manifold, such as deep autoencoders, and  alternative hyperreduction techniques, such as interpolation/least-squares fitting \cite{hirsch2024neural,kim2022fast} and collocation \cite{romor2023non,romor2023explicable}. The reader interested in the progress of this research front is urged to consult the literature surveys presented in these four references.
\end{remark}
}

\section{ Enhanced ECM with initial candidate points  }
\label{sec:Integration of Functions by Empirical Cubature}

 As   explained  in the preceding introductory section, the core ingredient of the proposed cubature rule with subspace-specific weights is a enhanced version  of the (discrete) Empirical Cubature Method (ECM) advocated by the authors in previous works \cite{hernandez2017dimensional,hernandez2020multiscale,hernandez2024cecm}.
 This enhanced  version  should be able to use as  input parameter a given set of candidate set of points.   In this section, we first outline the basic features of the original ECM when applied to global cubature problems, and  then elaborate on how to incorporate the proposed enhancement.   Readers not interested on the details of the argumentation may jump directly to the final   pseudo-code of this enhanced ECM, presented in Algorithm \ref{alg:ecm_global}.

\subsection{Global \own{c}ubature \own{p}roblem}

We tackle the problem of empirically devising a reduced integration rule for a vector-valued $\boldsymbol{\mu}$-parametric function
\begin{equation}
    \begin{aligned}
       \boldsymbol{a}: \ & \Omega \times \mathcal{P} \rightarrow \mathbb{R}^n  \\
        & (\boldsymbol{x}, \boldsymbol{\mu}) \mapsto (a_1, a_2, \dots, a_n) \ ,\\
    \end{aligned}
\end{equation}
\noindent defined over a domain $\Omega \subset \mathbb{R}^d$, with a parametric space $\mathcal{P}\subset \mathbb{R}^p$.  That is, we aim to find a set of integration points $\{ \bar{\boldsymbol{x}}_1, \dots, \bar{\boldsymbol{x}}_m\}$ with corresponding integration weights $\{\omega_1, \dots , \omega_m\}$, with $m$ as small as possible, such that
\begin{equation}
   \int_{\Omega} \boldsymbol{a} (\boldsymbol{x},\boldsymbol{\mu}) d \Omega \approx \sum_{g=1}^{m} \boldsymbol{a}( \bar{\boldsymbol{x}}_g,\boldsymbol{\mu}) \omega_g.
\end{equation}

To this end, we consider a finite element partition of the domain   $\bar{\Omega} = \bigcup_{e=1}^{N_{el}} \Omega^e $; for simplicity of exposition, the elements are assumed to be  isoparametric and  with the same number ($r$) of  Gauss points. If one considers $P$ samples of the parametric space:
\begin{equation}
    \{\boldsymbol{\mu}_i \}_{i=1}^{P} = \mathcal{P}^h \subset \mathcal{P} \ ,
\end{equation}
then, the integral of $\boldsymbol{a}$ over $\bar{\Omega}$ for all the samples in $\mathcal{P}^h$ is given by the elemental Gauss rule
\begin{equation}
    b_k = \sum_{e=1}^{N_{el}} \int_{\Omega^e} a_i (\boldsymbol{x}; \boldsymbol{\mu}_j) d\Omega^e = \sum_{e=1}^{N_{el}} \sum_{g=1}^{r} a_i (\bar{\boldsymbol{x}}_g^e, \boldsymbol{\mu}_j) W_g^e   \ \ \ k = (j-1) n+i   \ \ \ i = 1,2,\dots, n \ \ \ j = 1,2,\dots, P \ ,
    \label{eq: gauss quadrature a}
\end{equation}
\noindent where $\bar{\boldsymbol{x}}_g^e \in \Omega^e$ is the position of the $g$-th Gauss point of element $\Omega^e$, and $W_g^e$ is the product of the corresponding Gauss point weight and the Jacobian of the isoparametric transformation. The above expression can be written in matrix form as
\begin{equation}
    \begin{bmatrix}
        b_1 \\
        b_2 \\
        \vdots \\
        b_n \\
        b_{n+1} \\
        \vdots \\
        b_{nP}
    \end{bmatrix} =
        \begin{bmatrix}
        a_1(\bar{\boldsymbol{x}}^1_1 ; \mu_1 ) & a_1(\bar{\boldsymbol{x}}^1_2 ; \mu_1 ) & \dots &  a_1(\bar{\boldsymbol{x}}^1_r ; \mu_1 ) & a_1(\bar{\boldsymbol{x}}^2_1 ; \mu_1 )  & \dots &   a_1(\bar{\boldsymbol{x}}^{N_{el}}_r ; \mu_1 )\\
        a_2(\bar{\boldsymbol{x}}^1_1 ; \mu_1 ) &  a_2(\bar{\boldsymbol{x}}^1_2 ; \mu_1 ) & \dots &  a_2(\bar{\boldsymbol{x}}^1_r ; \mu_1 ) & a_2(\bar{\boldsymbol{x}}^2_1 ; \mu_1 )  & \dots &   a_2(\bar{\boldsymbol{x}}^{N_{el}}_r ; \mu_1 )\\
        \vdots &   \vdots      &\ddots &  \vdots & \vdots  & \ddots &   \vdots    \\
        a_n(\bar{\boldsymbol{x}}^1_1 ; \mu_1 ) & a_n(\bar{\boldsymbol{x}}^1_2 ; \mu_1 ) & \dots &  a_n(\bar{\boldsymbol{x}}^1_r ; \mu_1 ) & a_n(\bar{\boldsymbol{x}}^2_1 ; \mu_1 )  & \dots &   a_n(\bar{\boldsymbol{x}}^{N_{el}}_r ; \mu_1 )\\
        a_1(\bar{\boldsymbol{x}}^1_1 ; \mu_2 ) &  a_1(\bar{\boldsymbol{x}}^1_2 ; \mu_2 ) & \dots &  a_1(\bar{\boldsymbol{x}}^1_r ; \mu_2 ) & a_1(\bar{\boldsymbol{x}}^2_1 ; \mu_2 )  & \dots &   a_1(\bar{\boldsymbol{x}}^{N_{el}}_r ; \mu_2 )\\
        \vdots &   \vdots      &\ddots &  \vdots & \vdots  & \ddots &   \vdots    \\
        a_n(\bar{\boldsymbol{x}}^1_1 ; \mu_P ) & a_n(\bar{\boldsymbol{x}}^1_2 ; \mu_P ) & \dots &  a_n(\bar{\boldsymbol{x}}^1_r ; \mu_P ) & a_n(\bar{\boldsymbol{x}}^2_1 ; \mu_P )  & \dots &   a_n(\bar{\boldsymbol{x}}^{N_{el}}_r ; \mu_P )\\
        \end{bmatrix}
        \begin{bmatrix}
        W_1^1 \\
        W_2^1 \\
        \vdots \\
        W_r^1 \\
        W_1^2 \\
        \vdots \\
        W_r^{N_{el}}
        \end{bmatrix} \ ,
    \end{equation} \\
\noindent or in compact form
\begin{equation}
    \boldsymbol{b}^{FE} = \boldsymbol{A}^T \boldsymbol{W} \ ,
    \label{eq: integral function samples}
\end{equation}
\noindent where we will consider $\boldsymbol{b}^{FE} \in \mathbb{R}^{nP}$ as the exact integral over the discrete spatial and parameter spaces; $\boldsymbol{A} \in \mathbb{R}^{ M \times nP}  $, on the other hand, is the matrix containing the function evaluations at every Gauss point for the $P$ considered samples of the parameter space contained in $\mathcal{P}^h$ (  $M$ is the total number of Gauss points); finally the vector of weights  $\boldsymbol{W} \in \mathbb{R}^{M}$ contains the multiplication of the Gauss point weights and the Jacobians.

The standard,  global cubature problem \cite{hernandez2024cecm} consists in finding the smallest set of evaluation points, together with a  set of weights, such that the exact integral is approximated to a desired level of accuracy. This is posed as the following optimization problem:

\begin{equation}
    \begin{aligned}
        \min \quad & \norm{ \boldsymbol{\zeta} }_0\\
        \textrm{s.t.} \quad & \norm{ \boldsymbol{A}^T \boldsymbol{\zeta} - \boldsymbol{b}^{FE} }_2 \le \epsilon \norm{ \boldsymbol{b}^{FE} }_2    \\
        & \boldsymbol{\zeta} \succeq \boldsymbol{0} \ ,
    \end{aligned}
    \label{eq:optimization_problem_np_hard_1}
\end{equation}

\noindent where $\boldsymbol{\zeta} \in \mathbb{R}^{M}$ is a sparse vector, $\norm{\cdot}_0$ represents the so-called  $\ell_0$ pseudo norm (which counts the number of non-zero entries of its argument), $0 \le \epsilon \le 1$ is a user-defined tolerance, and the
symbol $\boldsymbol{\zeta} \succeq \zero$ represents that $\zeta_i \ge 0$ for all $i=1,2 \ldots M$.

\subsection{Standard ECM}
\label{sec: The Empirical Cubature Method}

Rather than operating directly over the matrix $\boldsymbol{A} \in \mathbb{R}^{ M \times nP}$,   the ECM operates on a basis matrix  $\boldsymbol{U} \in \mathbb{R}^{M \times \mP}$ whose columns are the  discrete representation of the basis function for the integrand functions. This basis matrix is obtained as the left singular vectors of a truncated SVD\footnote{ The left singular vectors of a standard SVD are columnwise orthogonal ($\U^T \U = \ident$). To obtain  $\W$-orthogonality, which is the one consistent with $L_2$ norm,   one must  follow the steps outlined in Appendix \ref{sec:appendix 2}.  Here, for simplicity of exposition, we use orthogonality in the Euclidean norm.   }

\begin{equation}
 [\boldsymbol{U} , \boldsymbol{\Sigma} , \boldsymbol{V} ] \leftarrow \texttt{SVD}(\boldsymbol{A}, \epsilon_{\text{\tiny SVD}}) \ ,
\label{eq:svd of A}
\end{equation}

\noindent where
\begin{equation}
    \boldsymbol{U}\in \mathbb{R}^{M \times \mP} \ \ , \ \ \boldsymbol{\Sigma}\in \mathbb{R}^{\mP \times \mP} \ \ , \ \ \boldsymbol{V}\in \mathbb{R}^{nP \times \mP} \ \ , \ \
    \norm{\boldsymbol{A} - \boldsymbol{U}\boldsymbol{\Sigma}\boldsymbol{V}^T}  \leq \epsilon_{\text{\tiny SVD}} \norm{\boldsymbol{A}} \ .
\end{equation}
\noindent The authors demonstrated in Ref. \cite{hernandez2024cecm}   that the truncation tolerance appearing in Eq. \ref{eq:optimization_problem_np_hard_1} and the truncation tolerance for the SVD in Eq. \ref{eq:svd of A} are of the same order, and therefore,   the integration error in the ECM may be  controlled by the SVD truncation tolerance. Likewise,  when the snapshot matrix is too large to fit into memory, the standard SVD can be replaced by the \emph{Sequential Randomized SVD} developed by the authors in the appendix of Ref. \cite{hernandez2024cecm}.

Having at our disposal the discrete representation of the basis function of the integrand, we can, similarly to Eq. \ref{eq: integral function samples}, define
\begin{equation}
    \b = {\boldsymbol{U}}^T \boldsymbol{W} \ ,
\end{equation}
\noindent  {where $\b \in \mathbb{R}^{\mP}$ is the vector containing the integral of the basis function, which for the empirical cubature problem will be considered the \textit{ground truth}. Moreover,  $\boldsymbol{W} \in \mathbb{R}^{M}$ is the same vector of weights appearing in Eq. \ref{eq: integral function samples}}. The ``global" optimization problem  in terms of the basis matrix $\U$ consists in finding the smallest subset of points and   weights that allow to \textit{exactly} recover the \emph{exact} or \emph{ground truth} integral of the basis functions, i.e.

\begin{equation}
    \begin{aligned}
        \min \quad & \norm{ \boldsymbol{\zeta} }_0\\
        \textrm{s.t.} \quad &  \boldsymbol{U}^T \boldsymbol{\zeta} - \b = \boldsymbol{0}    \\
        & \boldsymbol{\zeta} \succeq \boldsymbol{0} \ .
    \end{aligned}
    \label{eq:optimization_problem_np_hard_basis_function}
\end{equation}
\noindent It is well known that this  optimization problem involving the pseudo-norm  $\norm{\cdot}_0$  are computationally intractable (NP-hard) and therefore, recourse to either sub-optimal greedy heuristic or convexification is to be made \cite{boyd2004convex} .   The   ECM pertains to the category of  greedy algorithm,  and provides a non-optimal solution\footnote{ As demonstrated by authors in Ref. \cite{hernandez2024cecm}, it is possible within    the context of FE parametric problems to arrive at optimal solutions, but this requires considering the position of the points as continuous design variables.}, featuring as many points as columns in $\U$ . More specifically, as already pointed out in the introductory section,  the discrete ECM gives a solution to the problem of finding a set of $m$ rows $\Eind$ and positive weights $\boldsymbol{\omega}$ such that
\begin{equation}
\label{eq:fffff}
    \boldsymbol{U}_\Eind^T \boldsymbol{\omega}    =  \b,
\end{equation}
where $\U_{\Eind} = \U(\Eind,:) \in \RRn{m}{m}$.

\subsection{Candidate points as inputs}
\label{sec: non-uniqueness of ECM}

\begin{algorithm}[!ht]
\DontPrintSemicolon

\SetKwFunction{FMain}{ECM}
\SetKwProg{Fn}{Function}{}{}

\SetKwFunction{LSTONER}{LSTONER}
\SetKwFunction{UPHERM}{UPHERM}
\SetKwFunction{LENGTH}{length}

\Fn{ $[\Eind, \boldsymbol{\omega} ]$   =  \FMain{${\U}, \boldsymbol{W},  \boldsymbol{y_0}$ } }{
\KwData{$ {\U} \in \mathbb{R}^{ M \times m}$: where ${\U^T}{\U} = \boldsymbol{I}$; $\boldsymbol{W} \in \mathbb{R}^M_{+}$;  \ $\boldsymbol{y_0} \subseteq \{1,2 \ldots M \}$: initial candidates;  $\lambda \leftarrow 10$, threshold number of ineffective iterations}
\KwResult{$ \Eind \subseteq \{1,2,...,M \}$;
$\boldsymbol{\omega} \succ \boldsymbol{0}$ such that $\U(\Eind,:)^T \omegaB = \U^T \W$ and $\textrm{card}(\Eind \cap \y_0)$ is maximum. }
\vspace{2mm}
\eIf{
$\boldsymbol{y_0} = \emptyset $ \tcp{no initial candidate set given}
}{
$\boldsymbol{y} \leftarrow \{1, 2, \ldots M\}  $    \label{alg:01_no} \\
$\boldsymbol{y} \leftarrow \boldsymbol{y} \setminus \boldsymbol{h}$, where $\boldsymbol{h}=[h_1,h_2...]$ such that $\norm{{\U}(h_i,:)} \le \epsilon$  \tcp{Remove low norm points on the candidate set ($\epsilon \sim  10^{-6}$)}
} 
{
$\boldsymbol{y}' \leftarrow \{1, 2, \ldots M\} \setminus \boldsymbol{y_0}$\\
$\boldsymbol{y}' \leftarrow \boldsymbol{y}' \setminus \boldsymbol{h}$, where $\boldsymbol{h}=[h_1,h_2...]$ such that $\norm{{\U}(h_i,:)} \le \epsilon$  \tcp{Remove low norm points on the complement set ($\epsilon \sim  10^{-6}$)}
}
\vspace{2mm}
$\Eind \leftarrow \emptyset$; $\boldsymbol{b} \leftarrow {\U^T} \hspace{0.5mm} \boldsymbol{W}$;  $\boldsymbol{r} \leftarrow \boldsymbol{b}$;   $ \boldsymbol{\omega}  \leftarrow \emptyset$;  $\H \leftarrow \emptyset$ \label{alg: objective function definition} \tcp{Initializations}
$ \own{ \Theta}  \leftarrow 0$ \tcp{Failed iterations counter}
\vspace{2mm}
\While{  \LENGTH{$\Eind$}$ < p $   \textrm{\textbf{ AND }} \LENGTH($\boldsymbol{y}$)$> 0$  \label{line:while} }{
\If{
 $ \own{ \Theta}  >  \lambda$
}{
$\boldsymbol{y} \leftarrow \boldsymbol{y} \cup  \boldsymbol{y}'$  \label{line:enlarge} \tcp{Enlarge candidate set to include the complement set}
}
$i = \argmax_{i \in \boldsymbol{y}}$ $\boldsymbol{g}_{\boldsymbol{y}} \boldsymbol{r}$, where $\boldsymbol{g}_j = {\U}(j,:)/\norm{{\U}(j,:)}$ \label{line:pos} \tcp{Select the row most ``positively'' parallel to the residual}

\eIf{ $\Eind = \emptyset$}
    { $\H \leftarrow ({\U}(i,:) {\U}(i,:)^T)^{-1} $    \tcp{Inverse Hermittian matrix  (first iteration) }
      $\boldsymbol{\omega}  \leftarrow \H {\U}(i,:) \boldsymbol{b}$  \label{line:ls} \tcp{Weights computed through least-squares}
    } 
    {    [$ \boldsymbol{\omega}  ,\H$] $\leftarrow$ \LSTONER{$\boldsymbol{\omega} ,\H,{\U}({\Eind},:)^T,{\U}(i,:)^T,\boldsymbol{b}$   \label{line:r2} }   \tcp{Least-squares via rank-one update, see Algorithm 8 in Ref. \cite{hernandez2020multiscale}}
    } 
$\Eind \leftarrow \Eind \cup i$ ; $\; \;$   $\boldsymbol{y} \leftarrow \boldsymbol{y} \setminus i$; $\; \;$       \tcp{Move index $i$ from $\boldsymbol{y}$ to $\Eind$}

$\boldsymbol{n} \leftarrow $ Indexes such that $\boldsymbol{\omega} (\boldsymbol{n}) \prec \boldsymbol{0}$  \tcp{Identify negative weights}

\If{$\boldsymbol{n} \neq \emptyset$ \label{line:negative} }{

    $\boldsymbol{y} \leftarrow \boldsymbol{y} \cup \Eind(\boldsymbol{n})$; $\; \;$  $\Eind \leftarrow \Eind \setminus \Eind(\boldsymbol{n})$ ; $\; \;$    \tcp{Remove indexes negative weights}

    $\H$ $\leftarrow$ \UPHERM{$\H,\boldsymbol{n}$}  \label{line:r1} \tcp{  Update inverse Hermitian Matrix (via recursive, rank-one operations, see Algorithm 9 in Ref. \cite{hernandez2020multiscale} )}

    $\boldsymbol{\omega} = \H {\U}(\Eind,:) \boldsymbol{b}$  \tcp{Recalculate weights  \label{line:unc2} }

    } 
    \eIf{
    successful iteration \tcp{the size of $\Eind$ has increased w.r.t. the prior iteration}
    }{
    $  \own{ \Theta}   \leftarrow 0  $  \label{line:incre1}
    }{
    $ \own{ \Theta}  \leftarrow \own{ \Theta}  +   1  $  \label{line:incre2}
    }
    $\boldsymbol{r} \leftarrow \boldsymbol{b} - {\U}(\Eind,:)^T \boldsymbol{\omega} $ \label{line:res} \tcp{Update the residual}

}

}

\caption{  Empirical Cubature Method (enhanced version of the algorithm in Ref. \cite{hernandez2020multiscale} by incorporating an initial candidate set) }
\label{alg:ecm_global}
\end{algorithm}

In the standard ECM described in the foregoing,    the algorithm  searches for the reduced set of points  within the entire set of $M$ Gauss points of the mesh. However, as pointed out earlier,  for purposes of constructing the desired local ECM for subspace-specific weights cubature rules,  it is necessary to adapt this ECM so that it can cope with    situations in which the candidate points are but a subset   of all the Gauss points .

The pseudo-code of the    ECM adapted to this new scenario is outlined in Algorithm \ref{alg:ecm_global}.  The     inputs of   algorithm  are, on the one hand, the  basis matrix\footnote{As   pointed out  in the introductory section, to avoid ill-conditioned problems, the span of $\U$ should contain the vector space of constant functions; the procedure for guaranteeing  this condition is explained in Appendix \ref{sec:appendix 2}.} $\U$ and  the vector of full-order weights $\W$, and, on   the other  hand, the new input  $\y_0 \subseteq \{1,2 \ldots M \}$ ---the initial set of candidate indices. In what follows,    we limit ourselves to describe the modifications in  the algorithm related with this     new input. The reader is referred to Ref. \cite{hernandez2020multiscale}, appendix B, for   thorough details on   other ingredients  of the  algorithm ---such as the one-rank updates of Lines \ref{line:r1} and \ref{line:r2}.

 As noted in Refs. \cite{hernandez2017dimensional,hernandez2020multiscale}, the distinguishing feature of the ECM with respect to other methods based of the original nonnegative least-squares algorithm by Lawson and Hanson \cite{lawson1974solving}  (such as the ECSW \cite{farhat2014dimensional}) is that, in working with a orthogonal coefficient matrix $\U$, the  condition of positive weights need not be enforced during the iterations. As a consequence,   each iteration of the \emph{while} loop in Line \ref{line:while}  is in general an effective iteration,  in the sense that the unconstrained least-squares problem of Line \ref{line:r2} produces naturally a new cubature rule with an additional positive weight.

  Negative weights may   appear  during the iterations though,  caused by two different factors. The first factor is the appearance of vanishingly small weights  which   bounce between negative and positive values.
 This factor was identified in the original proposal  of Ref. \cite{hernandez2020multiscale}, and amended by introducing the conditional statement in  Line  \ref{line:negative},  in which indices with negative weights  are  removed, and the weights are updated again by solving the unconstrained least-squares of Line \ref{line:unc2}.

  The second factor concerns the new input $\y_0$.   It can be   readily demonstrated  that, for the algorithm to find a solution   within $\y_0$, the vector of integrals $\b = \U^T \W$ must pertain to the \emph{conical hull} (see e.g. \cite{hiriart1996convex}) defined by
\begin{equation}
\label{eq:83,sssss}
    \texttt{cone}( {\U(\y_0,:)}) := \left\{ \sum_{j=1}^{n} \alpha_j \U(\y_0(j),:)    \ \mid \  \alpha_j \ge 0   \right\} \
\end{equation}
(here $n$ is the cardinality of $\y_0$). A direct corollary of this condition is that  if $\y_0 = \{1,2 \ldots M \}$ (i.e.,    $\y_0$ is the entire set of   points),   then the  ECM will find invariably a solution,  since $\b$ is a positive linear combination of the rows of $\U$ (recall that $\W$ is by hypothesis  strictly positive).  However, for an arbitrary subset, the iterative algorithm in its original form is not guaranteed to arrive  at a feasible solution, the most obvious case being when the cardinality of $\y_0$ is smaller than the number of columns of $\U$ ($n < m$).   Empirically, we have found that lack of solutions manifests itself in the appearance of multiple negative weights. Since   negative weights and the associated indices are eliminated in Line \ref{line:negative}, when there is no solution   the algorithm ends up performing many ineffective iterations, and as  a  consequence the output $\Eind$ shrinks rather than expands as the algorithm progresses.   Accordingly, to identify when the provided initial set does not contain any feasible solution, we have   introduced a counter that tracks the number of   ineffective or failed iterations (Line \ref{line:incre2}). When this number  exceeds a user-prescribed threshold (we use $\lambda = 10$ in all the examples presented in the paper), the initial set of candidate indices is enlarged to encompass the remaining indices (Line \ref{line:enlarge}). With this simple  heuristic ---enlarging the set of candidate indices in case of convergence failure---, the ECM  is always guaranteed to provide a solution to the problem: at best, it will find indices which are indeed a subset of the initial candidate set ($\Eind \subseteq \y_0$); at worst, if there is no solution within $\y_0$,  it will try to  maximize the elements of $\Eind$ that are in $\y_0$.

\subsection{Graphical interpretation}

\begin{figure}[!ht]
  \centering
   \input{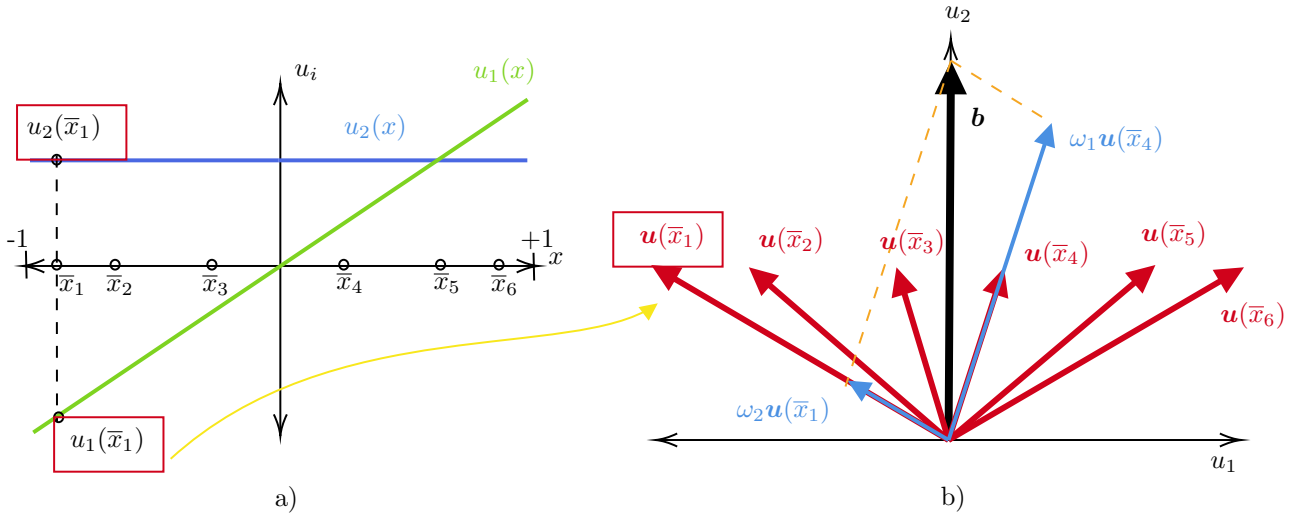}
   \caption{Performance of the Empirical Cubature Method (ECM) for the case of the    $u_1 = \sqrt{3/2} x$ and  $u_2 = \sqrt{1/2}$   in the interval $\Omega = [-1,1]$ with $M= 6$ Gauss integration points. Note that $\u(\xG{i}) = \U(i,:)$. }
     \label{fig:Figure2PB}
\end{figure}

The toy problem presented in Figure \ref{fig:Figure2PB} may help to clarify how the enhanced ECM explained above actually works.  The problem  consists in evaluating the integral on $\Omega = [-1,1]$ of two orthogonal polynomial functions, $u_1 = \sqrt{3/2} x$ and $u_2 = \sqrt{1/2}$,  by using  $M= 6$ Gauss points. The functions and locations of the points are represented in Figure \ref{fig:Figure2PB}.a.    The discrete quadrature problem with positive weights in this case  thus boils down   to selecting $m = 2$ points out of $M = 6$ possible candidates, which implies that  there is a total of   ${\ngausT \choose \pNMOD} = {6 \choose 2} = 15$ combinations of pairs that are  potential  solutions of the problem. However, not all of them will yield positive weights.  This can be appreciated in Figure  \ref{fig:Figure2PB}.b, where we represent  the values of the functions at the 6 points in the $u_1-u_2$ plane (note that $\U(i,j) = u_j(\xG{i})$) along with the vector of exact integrals $\b = [0,\sqrt{2}]^T $.  It can be readily seen that $\{\xG{1},\xG{2}\}$ is not a feasible solution because, as stated previously, see Eq. \ref{eq:83,sssss}, $\b$ does not pertain to the conical hull  of $\u(\xG{1})$ and $\u(\xG{2})$ ---or, in other words, it cannot be expressed as  a positive linear combination of $\u(\xG{1})$ and $\u(\xG{2})$. By the same token, $\{\xG{1},\xG{4}\}$ and $\{\xG{3},\xG{6}\}$ are feasible solutions  because $\b$  does     belong  to the cone positively spanned by $\{\u(\xG{1}),\u(\xG{4})\}$ and $\{\u(\xG{3}),\u(\xG{6})\}$, respectively.

To understand the logic of why Algorithm  \ref{alg:ecm_global} chooses one feasible solution or another depending on the initial set, let us consider first the case in which all points are included in the initial candidate  set  ($\y_0 = \{1,2,\ldots 6\}$).  In the first iteration,  we have that    $\r = \b$, that is, the residual vector is equal to the vector of exact  integrals $\b$ (Line \ref{alg: objective function definition}). Then, in Line \ref{line:pos}, the algorithm seeks the index $i$ for which $\U(i,:)$ is more   positively parallel  to the current residual (which is $\b$).     As can be observed in Figure \ref{fig:Figure2PB}.b, this vector  is      $\u(\xG{4}) =  \U(4,:)$, so  for the first iteration we have that  $\Eind =  \{ 4\}$   (because of symmetry, the algorithm may have chosen $\Eind =  \{ 3\}$ as well). Then the algorithm computes the weight for this point through least-squares (Line \ref{line:ls}), and updates the residual in Line \ref{line:res}. Although not plotted in Figure \ref{fig:Figure2PB}.b, it can be inferred that this updated residual will be most positively parallel to $\u(\xG{4})$; thus, the final set of indices will be $\Eind = \{4,1\}$.

Now suppose    we remove index $i=1$ from the initial set, i.e., $\y_0 = \{2,3,4,5,6\}$; it follows, by the logic explained above, that the the final solution will be $\Eind = \{4,2\}$; if index $i=2$ is also removed, then $\Eind = \{4,3\}$. However if $\y_0 = \{4,5,6\}$, the most positively parallel vector in the second iteration will be $\u(\xG{5})$, and  $\b$ does not fall into the conic hull of  $\{\u(\xG{4}),  \u(\xG{5}) \}$. Consequently, the least-squares projection will produce negative weights. After a number of iterations with negative weights greater than $\lambda$, the algorithm will end up enlarging the set of candidates (Line \ref{line:enlarge}), and returning  a feasible solution ---which in  this particular case will be  $\Eind = \{4,1\}$.

\jahoHIDE{

Taking a subset of these candidates, \( \boldsymbol{y_0} \subseteq \boldsymbol{y} \), we can define the set of conic combinations (conical hull) of \( \boldsymbol{y_0} \) as:

\begin{equation}
    \texttt{cone}(\boldsymbol{y_0}) := \left\{ \sum_{j=1}^{m} \alpha_j \boldsymbol{\lambda}_j   \ \mid \  \boldsymbol{\lambda}_j \in \boldsymbol{y}_0, \alpha_j \in \mathbb{R}_+ ,  m \in \mathbb{N} \right\} \ .
\end{equation}

It follows that a solution can be found when $ \b \in \texttt{cone}(\boldsymbol{y_0})$. Indeed, a solution can be obtained by any conic combination of the rows of matrix $\boldsymbol{U}$ containing the vector $\b$. This fact will be exploited for proposing the local empirical cubature method in the following sections.

In line with the previous discussion on the non-uniqueness of the ECM solution in Sec. \ref{sec: non-uniqueness of ECM}, suppose \( \boldsymbol{y} \) encapsulates all points stored row-wise in a generic matrix \( \boldsymbol{U}^{(i)} \). Let \( \boldsymbol{y_0} \subseteq  \boldsymbol{y} \) be the set of initial candidate points. A solution exists if the ground truth integral lies within the convex hull of \( \boldsymbol{y_0} \), symbolically represented as \( {\boldsymbol{U}^{(i)}}^T \boldsymbol{W} = \b^{(i)} \ni \texttt{cone}(\boldsymbol{y_0}) \). If this condition is not met, the ECM algorithm considers as many points from the candidate set as possible and then turns to the complementary set \( \boldsymbol{y}' = \boldsymbol{y} \setminus \boldsymbol{y_0} \) for additional candidates.

For illustrative purposes, consider Fig. \ref{fig:Figure2PB}. If we consider an initial candidate set  \( \boldsymbol{y_0} = \revone{\{\boldsymbol{u}(\bar{x_1}),\boldsymbol{u}(\bar{x_2}),\boldsymbol{u}(\bar{x_3})\} }\), it is easy to observe that no conic combination of these vectors contains \( \b \). Thus, the candidate set must be expanded to integrate the complementary set, which for this case would be given as \( \boldsymbol{y}' = \revone{\{ \boldsymbol{u}(\bar{x_1}) ,\dots, \boldsymbol{u}(\bar{x_6}) \} \setminus \boldsymbol{y_0} = \{ \boldsymbol{u}(\bar{x_4}), \boldsymbol{u}(\bar{x_5}), \boldsymbol{u}(\bar{x_6}) \}} \).

In this regard, it should be stressed that, as argued by the authors in Ref. \cite{hernandez2024cecm}, the solution of the discrete global cubature method is not unique

the solution of the discrete cubature problem is not unique, but grows

However, the ability of the algorithm to find the solution within this set of candidate points will depend on the particular subset

convergence of the greedy algorithm underlying the ECM is only guaranteed when all the points are included;  if not,   convergence will depend on the particular subset of candidate points. To overcome,

However, as we argue in the sequel, one may search for this solution set withing a smaller subset of points and still find  a solution. To illustrate this idea,
Upon examining the graphical interpretation of the ECM in Fig. \ref{fig:Figure2PB}, the non-uniqueness becomes evident. The figure depicts the integration of a basis function

defined in Eq. \ref{eq:optimization_problem_np_hard_basis_function} admits multiple solutions. The solution provided by the ECM represents just one among a combinatorially vast set of possibilities. We delve deeper into this non-uniqueness aspect in the subsequent section.

Upon examining the graphical interpretation of the ECM in Fig. \ref{fig:Figure2PB}, the non-uniqueness becomes evident. The figure depicts the integration of a basis function

\begin{equation}
    \begin{aligned}
       \boldsymbol{u}: \ & [0,1] \rightarrow \mathbb{R}^2  \\
        & x  \mapsto (u_1, u_2) \ ,\\
    \end{aligned}
\end{equation}

\noindent using a single finite element $N_{\text{el}}=1$ with six Gauss points $r=6$. The matrix containing the discrete representation of these basis function is then $\boldsymbol{U}\in \mathbb{R}^{6 \times 2}$. The 2-dimensional vectors contained row-wise in $\boldsymbol{U}$ are represented by red arrows in the $u_1u_2$ plane in Fig. \ref{fig:Figure2PB} b). The problem to be solved can be boiled down to choosing any two directions and corresponding weights such that vector $\b = \boldsymbol{U}^T \boldsymbol{W}$ is exactly represented. The ECM selects vectors 4 and 1. However, we note that selecting any vector to the left and any other vector to the right of $\b$ would allow to compute positive weights such that the vector $\b$ is exactly approximated.}

\jahoHIDE{
Consider now the case or a general matrix $\boldsymbol{U} \in \mathbb{R}^{M \times \mP}$. We define the complete candidates for the ECM as the set comprising the \(\mP\)-dimensional vectors contained row-wise within \( \boldsymbol{U} \), which can be represented as

\begin{equation}
\boldsymbol{y} := \{ \boldsymbol{U}[1,:], \boldsymbol{U}[2,:], \ldots, \boldsymbol{U}[rN_{\text{el}},:]\} \ .
\end{equation}

Taking a subset of these candidates, \( \boldsymbol{y_0} \subseteq \boldsymbol{y} \), we can define the set of conic combinations (conical hull) of \( \boldsymbol{y_0} \) as:

\begin{equation}
    \texttt{cone}(\boldsymbol{y_0}) := \left\{ \sum_{j=1}^{m} \alpha_j \boldsymbol{\lambda}_j   \ \mid \  \boldsymbol{\lambda}_j \in \boldsymbol{y}_0, \alpha_j \in \mathbb{R}_+ ,  m \in \mathbb{N} \right\} \ .
\end{equation}

It follows that a solution can be found when $ \b \in \texttt{cone}(\boldsymbol{y_0})$. Indeed, a solution can be obtained by any conic combination of the rows of matrix $\boldsymbol{U}$ containing the vector $\b$. This fact will be exploited for proposing the local empirical cubature method in the following sections.

In line with the previous discussion on the non-uniqueness of the ECM solution in Sec. \ref{sec: non-uniqueness of ECM}, suppose \( \boldsymbol{y} \) encapsulates all points stored row-wise in a generic matrix \( \boldsymbol{U}^{(i)} \). Let \( \boldsymbol{y_0} \subseteq  \boldsymbol{y} \) be the set of initial candidate points. A solution exists if the ground truth integral lies within the convex hull of \( \boldsymbol{y_0} \), symbolically represented as \( {\boldsymbol{U}^{(i)}}^T \boldsymbol{W} = \b^{(i)} \ni \texttt{cone}(\boldsymbol{y_0}) \). If this condition is not met, the ECM algorithm considers as many points from the candidate set as possible and then turns to the complementary set \( \boldsymbol{y}' = \boldsymbol{y} \setminus \boldsymbol{y_0} \) for additional candidates.

For illustrative purposes, consider Fig. \ref{fig:Figure2PB}. If we consider an initial candidate set  \( \boldsymbol{y_0} = \revone{\{\boldsymbol{u}(\bar{x_1}),\boldsymbol{u}(\bar{x_2}),\boldsymbol{u}(\bar{x_3})\} }\), it is easy to observe that no conic combination of these vectors contains \( \b \). Thus, the candidate set must be expanded to integrate the complementary set, which for this case would be given as \( \boldsymbol{y}' = \revone{\{ \boldsymbol{u}(\bar{x_1}) ,\dots, \boldsymbol{u}(\bar{x_6}) \} \setminus \boldsymbol{y_0} = \{ \boldsymbol{u}(\bar{x_4}), \boldsymbol{u}(\bar{x_5}), \boldsymbol{u}(\bar{x_6}) \}} \).}

\section{Cubature problem with multiple subspaces}

Attention is confined now to the problem of deriving a subspace-adaptive weights cubature rule for a given collection of  subspaces of integrable functions.  Let us consider a set of $k$ vector-valued $\boldsymbol{\mu}$-parametric functions
\begin{equation}
    \begin{aligned}
       \boldsymbol{a}^{(i)}: \ & \Omega \times \mathcal{P} \rightarrow \mathbb{R}^{n_i} \ ,  \ \ \  \ (i = 1,2, \dots k) \\
       & (\boldsymbol{x}, \boldsymbol{\mu}) \mapsto \left(a_1^{(i)} , a_2^{(i)} \dots, a_{n_i}^{(i)}  \right)\ , \ \ \ \ (i = 1,2, \dots k) \ . 
    \end{aligned}
\end{equation}
Furthermore,  for each of the functions $ \boldsymbol{a}^{(i)}$,   consider $P_i$ samples of the parametric space  such that
\begin{equation}
    \{\boldsymbol{\mu}_m \}_{m=1}^{P_i} = \mathcal{P}_i^h \ ,  \ \ \ \ \bigcup_{i=1}^{k} \mathcal{P}_i^h  \subset \mathcal{P} \ . 
\end{equation}
This allows to define $k$ matrices containing the evaluations of the $i-$th function for all Gauss points and  for each point in the corresponding discrete parametric space $\mathcal{P}_i^h$  as follows
\begin{equation}
   \boldsymbol{A}^{(i)} \in  \mathbb{R}^{ M \times n_iP_i}  \ \ \ \ , \ (i = 1,2, \dots k) \ .
    \label{eq: A_i}
\end{equation}
Moreover, for each function  {we can identify} the vectors containing what we consider their exact integral as
\begin{equation}
     \boldsymbol{b}^{(i)}_{FE} = {\boldsymbol{A}^{(i)}}^T \boldsymbol{W} \ \ \ \ , \ (i = 1,2, \dots k) \ .
    \label{eq: objective local}
\end{equation}
 {We note that although we employ the superscript $\boldsymbol{\bullet}^{(i)}$ to refer to the distinct subspace of functions and their matrices of discrete evaluations, the weights vector $\boldsymbol{W} \in \mathbb{R}^{M}$ has no superscript. This is because the weights vector is exclusively associated with the discretization of the spatial domain and remains the same for all subspaces. 

Similarly to  the case of the standard, global cubature problem of Section \ref{sec: The Empirical Cubature Method},
  we will cast  the optimization problem with respect to a set of orthogonal   basis matrices $\U^{(1)},\U^{(2)} \ldots \U^{(k)}$. These orthogonal basis matrices are obtained via   a truncated SVD as in Eq. \ref{eq:svd of A}:
\begin{equation}
     \mathbb{R}^{ M \times n_iP_i} \ni \boldsymbol{A}^{(i)} \xrightarrow[\text{}]{\text{ SVD}} \boldsymbol{U}^{(i)} \in  \mathbb{R}^{ M \times \mP_i }  \ \ \ \ , \ (i = 1,2, \dots k) \ .
    \label{eq: U_i}
\end{equation}
The cubature problem we wish to solve is thus the following:  given  $\U^{(1)},\U^{(2)} \ldots \U^{(k)}$ and  $\W$, find the smallest set of  $m$ points $\Eind \subseteq \{1,2 \ldots M\}$ (the same set for  all the subspaces) and the associated nonnegative weighs $\omegaB^{(1)},\omegaB^{(2)} \ldots  \omegaB^{(k)}$ such that
\begin{equation}
 \U^{(i)^T}_{\Eind} \omegaB^{(i)}  = {\U^{(i)^T}} \W
\end{equation}
where $\U^{(i)}_{\Eind} = \U^{(i)}(\Eind,:)$.

\subsection{Subspace-adaptive weights ECM  }
\label{sec: The Local Empirical Cubature Method}

\begin{algorithm}[!ht]
    \DontPrintSemicolon

    \SetKwFunction{FMain}{SAW\_ECM}
    \SetKwProg{Fn}{Function}{}{}

    \SetKwFunction{LSTONER}{LSTONER}
    \SetKwFunction{UPHERM}{UPHERM}
    \SetKwFunction{LENGTH}{length}
    \SetKwFunction{ECM}{ECM}

    \Fn{ $[\Eind, \{\boldsymbol{\omega}^{(i)}\}_{i=1}^k ]$   $\leftarrow$  \FMain{$\{\boldsymbol{A}^{(i)}\}_{i=1}^k$ , $\boldsymbol{W}$, $\epsilon_{\text{\tiny SVD}}$ } }{
    \KwData{ $\mathbb{R}^{ M \times n_iP_i} \ni \{ \boldsymbol{A}^{(i)} \}_{i=1}^k$: Set of $k$ matrices containing the sampling, in the discrete spatial and parametric spaces, of the function to integrate; $\mathbb{R} ^{M} \ni\boldsymbol{W}$: vector containing the multiplication of the Jacobian of the isoparametric transformations multiplied by its corresponding Gauss point weights (all positive); $0 \le \epsilon_{\text{\tiny SVD}} \le 1$: truncation tolerance for the SVDs}
    \KwResult{$\Eind$ : Set of Gauss point indices ; $\{\boldsymbol{\omega}^{(i)}\}_{i=1}^k$ : $k$ sets of weights such that $\norm{{\boldsymbol{A}_\Eind^{(i)}}^T \boldsymbol{\omega}^{(i)} -  \boldsymbol{b}^{(i)}}_2  \sim \epsilon_{\text{\tiny SVD}} \norm{\boldsymbol{b}^{(i)} }_2$, where $\boldsymbol{b}^{(i)} = {\boldsymbol{A}^{(i)}}^T \boldsymbol{W} $ , for $ \ (i = 1,2, \dots k)$ }
    \For(){i=1 to k \label{line:saw1}}{
    $[ \boldsymbol{U}^{(i)}, \boldsymbol{\Sigma}^{(i)}, \boldsymbol{V}^{(i)} ] \leftarrow \texttt{SVD}(  \boldsymbol{A}^{(i)}  , \epsilon_{\text{\tiny SVD}}$ ) \tcp{Truncated SVD of each sample matrix. }
    $\U^{(i)}  \leftarrow$  \texttt{augment\_matrix}($\U^{(i)},\W$) \label{line:augment} \tcp{Augment matrix with an additional column to avoid ill-posedness, see Appendix \ref{sec:ecmp}}
    }
    indices $\leftarrow$ \texttt{sort\_index}($\{\boldsymbol{U}^{(i)}\}_{i=1}^k$) \label{line:saw5}  \tcp{Sort indices $1,2 \ldots k$ according to some specified criterion }
    $\y_0 \leftarrow  \emptyset$ \\
    \For(){ $j$ in \texttt{indices}  \label{line:saw2}}{
    $[\Eind^{(j)}, \w^{(j)} ] \leftarrow   $ \ECM{${\boldsymbol{U}^{(j)}}$, $\boldsymbol{W}$ , $\y_0$}  \label{alg:2ECM} \tcp{This is the enhanced ECM described in Alg. \ref{alg:ecm_global}. It finds $\Eind^{(j)}$ and $\w^{(j)}>0$ such that $\U^{(j)}(\Eind^{(j)},:)^T \w^{(j)} = {\U^{(j)}}^T \W$, and at the same time maximizes the number of elements in $\Eind^{(j)}$   pertaining  to $\y_0$  }

    $  \y_0 \leftarrow  \y_0   \cup \Eind^{(j)}$   \label{line:saw07}  \\
    }
    $\Eind \leftarrow \y_0$

     \For(){ $j=1$  \texttt{to}   $k$  \label{line:saw3}}{
     $\omegaB^{(j)} \leftarrow  \texttt{assemb\_weights}(\w^{(j)},\Eind^{(j)},\Eind)$  \label{line:saw4} \tcp{ Initialize $\omegaB^j$ as an $m$-zeros vector and assign $\w^j$ to the positions in $\omegaB^j$  occupied by $\Eind^{(j)}$ within $\Eind$  (here    $m = \texttt{length}(\Eind)$)}
    }

    }

    \caption{Subspace-Adaptive Weights ECM (SAW-ECM)}
    \label{alg:ecm_kratos_ECM_local_pod}
    \end{algorithm}

We  will   develop two distinct, novel solution strategies for the preceding cubature problem: one based on   the enhanced   ECM discussed in Section \ref{sec:Integration of Functions by Empirical Cubature}, and another one based on linear programming. We address here the  strategy based on the ECM, which    we will refer to  as either   the  \emph{ Subspace-Adaptive Weights ECM} (  SAW-ECM) or simply the \emph{Local ECM}, and whose pseudo-code is described in  Algorithm    \ref{alg:ecm_kratos_ECM_local_pod}.      The inputs of the method are the sample matrices $\A^{(1)},\A^{(2)}, \ldots \A^{(k)}$, the vector of weights $\W$ and the truncation tolerance for the SVDs ($\epsilon_{\tiny SVD}$). The first part of the algorithm (Line \ref{line:saw1})  computes the basis matrices $\U^{(1)},\U^{(2)} \ldots \U^{(k)} $  for the column space of the sample matrices ($\U^i \in \RRn{M}{n_i}$).
The loop  in Line \ref{line:saw2}, on the other hand, recurrently exploits the        enhancement introduced  in the ECM of Algorithm \ref{alg:ecm_kratos_ECM_local_pod}, whereby the method accepts as  input a set of candidate indices $\y_0$, and returns a set of indices $\Eind^{(j)}$ such that the     cardinality of the intersection of  $\y_0$ and   $\Eind^{(j)}$ is maximum (in other words, the algorithm tries to use as many elements in $\y_0$ as possible to construct $\Eind^{(j)}$).   The final set of indexes $\Eind$ is the union of the subsets produced in each call of the ECM. As for the weights, by virtue of the properties of the ECM, each output weights $\w^{(j)}$ has $m_j$ positive entries (as many as columns in $\U^j$). However, the problem demands that the         output weigths $\omegaB^{(j)}$ ($j=1,2 \ldots k$)    have all $m$ components ($m = \textrm{card}(\Eind)$). To ensure this,   Line \ref{line:saw4} initializes each   $\omegaB^j$ to a $m$-zeros vector, and then assigns $\w^j$ to the positions in $\omegaB^j$  corresponding to the positions occupied  by $\Eind^{(j)}$ in $\Eind$.

\revtwo{
\begin{remark}
   In the first iteration, when searching for the integration points of the first subspace $\U^{(1)}$, the SAW-ECM uses as candidate points the entire set of  Gauss points (this is indicated in Line \ref{line:saw07} of Algorithm \ref{alg:ecm_kratos_ECM_local_pod} and Line \ref{alg:01_no} of Algorithm \ref{alg:ecm_global}).  For the second iteration, the set of points calculated for $\U^{(1)}$  serves as the candidate set for the second basis matrix $\U^{(2)}$, and so on until  visiting all the basis matrices. It follows then that the final set of integration points depends on how the sequence of basis matrices is visited: different permutations of   $\U^{(1)},\U^{(2)}, \ldots \U^{(k)}$ will produce different final sets of integration points. This sensitivity to the order in which the basis matrices are processed will be   studied in the numerical example presented in Section \ref{sec:caseMAX}.
\end{remark}
}

\subsection{ Linear \own{p}rogramming-based approach}
\label{sec:LP}

 The second method   proposed herein for dealing with the cubature problem with multiple subspaces is based  on
 the same strategy underlying the Empirical Quadrature Method of Patera et al. \cite{patera2017lp,yano2019lp}  for the standard  cubature problem. This strategy consists in, first,  formulating the  cubature problem as a sparsification problem,  and then convexify it by replacing the $\ell_0$ pseudo-norm that characterizes sparsification problems   by the  $\ell_1$ norm ---it is well-known that objective functions defined in terms of this norm also promote sparsity  \cite{bruckstein2009sparse, elhamifar2013sparse}.

 \begin{figure}[!ht]
    \centering
    \includegraphics[width=.85\linewidth]{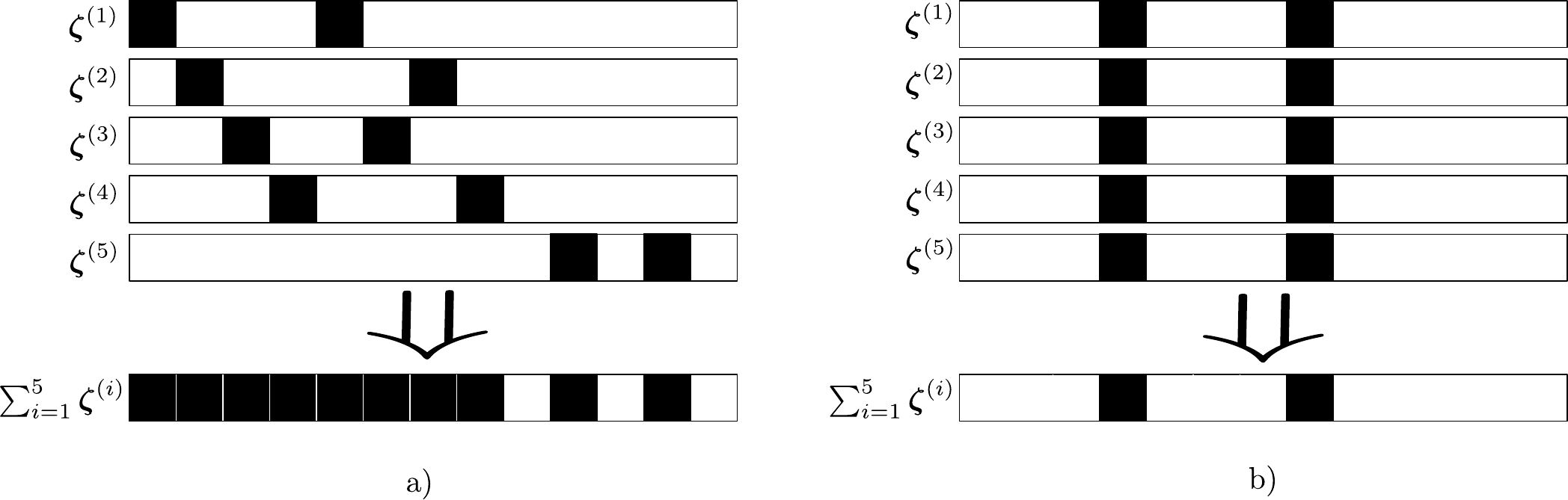}
    \caption{Sparsity plots for two feasible sets of sparse vectors $\boldsymbol{\zeta}^{(i)}$ for a case with $M= 13$   points,   $k=5$ subspaces. For both cases, each   $\boldsymbol{\zeta}^{(i)}$  ($i=1,2 \ldots 5$) contains 2 nonzero entries. However, while in one case (Figure a) the non-zero entries are spread among the 13 indices, in the other case (Figure b) they lie on the same indices for the five subspaces. This fact can be quantified by counting the nonzero entries of the sum of the weights (which is the objective function of the optimization problem \ref{eq:local l0 problem} ): a)  $\norm{\sum_{i=1}^{5} \boldsymbol{\zeta}^{(i)} }_0 = 10$; and  b) $\norm{\sum_{i=1}^{5} \boldsymbol{\zeta}^{(i)} }_0 = 2$.  }
    \label{fig: sparsity concept}
\end{figure}

The formulation of the cubature problem as a sparsification problem requires first     to consider the design variables as weight vectors   $\zetaI{1}, \zetaI{2} \ldots \zetaI{k}$ of the same size as the FE weight vector  $\W$, and, secondly, to use as  objective function  the total number of integration points $m  =\textrm{card}(\Eind)$ (as done in the global, standard cubature problem  \ref{eq:optimization_problem_np_hard_basis_function}).  Since the weights are forced to be nonnegative,   it follows that  this number is simply equal to the number of positive entries of the sum of all the weights, i.e., $ m = \norm{ \sum_{i=1}^{k} \boldsymbol{\zeta}^{(i)} }_0$ ---this idea is   illustrated in the sparsity plots displayed in Figure  \ref{fig: sparsity concept}.   With these considerations in mind, the  sparsification problem  for the case of multiple subspaces   may be posed as follows:   given $\Ui{1},\Ui{2} \ldots \Ui{k}$  ($\Ui{i} \in \RRn{M}{m_i}$) and $\bI{i} = {\Ui{i}}^T \W$ ($i=1,2  \ldots k$), find vector weights $\zetaI{1}, \zetaI{2} \ldots \zetaI{k}$    such that:
\begin{equation}\label{eq:local l0 problem}
    \begin{aligned}
        \min \limits_{\boldsymbol{\zeta}^{(1)},\boldsymbol{\zeta}^{(2)}, \dots , \boldsymbol{\zeta}^{(k)} } \quad & \norm{ \sum_{i=1}^{k} \boldsymbol{\zeta}^{(i)} }_0\\
        \textrm{s.t.} \quad &   {\boldsymbol{U}^{(i)}}^T \boldsymbol{\zeta}^{(i)} - \boldsymbol{b}^{(i)}  = \boldsymbol{0} \hspace{3mm} , \hspace{2mm} (i = 1, 2, \dots, k) \\
        & \boldsymbol{\zeta}^{(i)} \succeq \boldsymbol{0} \hspace{3mm} , \hspace{2mm} (i = 1, 2, \dots, k) \ .
    \end{aligned}
\end{equation}

As pointed out earlier, the convexification of the preceding problem is done by replacing the $\ell_0$ in the objective function by  the $\ell_1$-norm; furthermore, since the weights are positive, we have that
\begin{equation}
    \norm{ \sum_{i=1}^{k} \boldsymbol{\zeta}^{(i)} }_1 =   \sum_{i=1}^{k} \sum_{j=1}^{M}  \zeta_j^{(i)},
\end{equation}
and therefore the   problem, which was already linear in the constraints, becomes also  linear in the objective function:
\begin{equation}\label{eq:original l1 problem}
    \begin{aligned}
        \min \limits_{\boldsymbol{\zeta}^{(1)},\boldsymbol{\zeta}^{(2)}, \dots , \boldsymbol{\zeta}^{(k)} } \quad &  \sum_{i=1}^{k} \sum_{j=1}^{M} \zeta_j^{(i)} \\
        \textrm{s.t.} \quad &   {\boldsymbol{U}^{(i)}}^T \boldsymbol{\zeta}^{(i)} - \boldsymbol{b}^{(i)}  = \boldsymbol{0} \hspace{5mm} (i = 1, 2, \dots, k) \\
        & \boldsymbol{\zeta}_i \succeq \boldsymbol{0} \hspace{5mm} (i = 1, 2, \dots, k) \ ,
    \end{aligned}
\end{equation}
and, thus, amenable to be solved by linear programming techniques.  More specifically, it can be cast in the standard form \cite{Nocedal1999} in terms of a single design variable $\boldsymbol{\tilde{\zeta}}$ as follows:
\begin{equation}\label{eq:linear programming problem}
    \begin{aligned}
        \min \limits_{\boldsymbol{\tilde{\zeta}}} \quad & \mathbbm{1}^T \tilde{\boldsymbol{\zeta}}\\
        \textrm{s.t.} \quad &   \tilde{\boldsymbol{U}}^T \tilde{\boldsymbol{\zeta}} - \tilde{\boldsymbol{b}}  = \boldsymbol{0} \\
        & \tilde{\boldsymbol{\zeta}} \succeq \boldsymbol{0} \ ,
    \end{aligned}
\end{equation}
Here,   $\mathbbm{1}$ is an all-ones vector of length $k M$, whereas
\begin{equation}
    \begin{aligned}
     \tilde{\boldsymbol{\zeta}}:=
    \begin{bmatrix}
    \boldsymbol{\zeta}^{(1)} \\
    \boldsymbol{\zeta}^{(2)} \\
    \vdots \\
    \boldsymbol{\zeta}^{(k)} \\
    \end{bmatrix}
    & \hspace{5mm}
    \tilde{\boldsymbol{b}} :=
    \begin{bmatrix}
    \boldsymbol{b}^{(1)} \\
    \boldsymbol{b}^{(2)} \\
    \vdots \\
    \boldsymbol{b}^{(k)} \\
    \end{bmatrix}
    & \hspace{5mm}
    \tilde{\boldsymbol{U}} := \texttt{diag}\left({\boldsymbol{U}^{(1)}}^T, {\boldsymbol{U}^{(2)}}^T, \dots , {\boldsymbol{U}^{(k)}}^T  \right)   \ .
    \end{aligned}
    \label{eq: lp vectors definition}
\end{equation}
Once the sparse weights have been computed, the set of solution indices $\Eind$ is determined as the indices with positive entries in $\sum_{i=1}^{k} \boldsymbol{\zeta}^{(i)} $. Likewise, the solution weights are obtained from the sparse vectors by simply making $\omegaB^{(i)} = \zetaI{i}(\Eind)$ ($i=1,2 \ldots k$). Hereafter  we will symbolyze all these operations  as
\begin{equation}
    \left(\Eind,   \{\boldsymbol{\omega}^{(i)}\}_{i=1}^k  \right)\leftarrow \texttt{LP} \left( \tilde{\boldsymbol{U}}, \mathbbm{1} , \tilde{\boldsymbol{b}},  \tilde{\boldsymbol{\zeta}} \succeq \boldsymbol{0} \right) \ .
\end{equation}

\subsection{Remark on computational cost and efficiency }
\label{sec:ramk}
The technique described above may prove appealing (in comparison with the SAW-ECM)  because it  can leverage well-established methods and algorithms of linear programming (see e.g. Ref. \cite{Nocedal1999}).  However, it should be noted that      the size of the design variable $\tilde{\zetaB}$ in this approach is equal to the number of points of the  finite element mesh times the number of basis matrices; this fact may   render it prohibitively costly for relatively large problems.  By constrast,  the SAW-ECM, because of its ``greedy'' character,    only needs to process one basis matrix at a time; besides, the   computational cost of each  call of the  ECM in Line \ref{line:saw2} of  Algorithm \ref{alg:ecm_kratos_ECM_local_pod} depends only on the square of the number of columns of the basis matrix (this is demonstrated  in Ref. \cite{hernandez2020multiscale}, appendix A), but not on the number of Gauss points.

This remark aside, the focus of the ensuing numerical assessment is not on computational cost per se, but on which   of  the two approaches (SAW-ECM or Linear Programming) provides the smallest number of points for the same subspaces of functions.  In this regard, we note that the number of integration points for both the SAW-ECM  and the LP-based method is bounded below by the maximum of the number of columns of the basis matrices (i.e.,  $m \ge m_{max}$, where $m_{max} = \textrm{max}(m_i)$, $i=1,2 \ldots k$).   Likewise, it may be argued that  an upper bound for the number of integration points when using the SAW-ECM   is the dimension, hereafter denoted by $m_{all}$, of the alluded to earlier global subspace ---the sum of the spans of all the basis matrices. A basis matrix for this subspace may be obtained by applying the SVD to the concatenation of all the basis matrices $\U =[\U^{(1)},\U^{(2)} \ldots \U^{(k)}]$. In summary
\begin{equation}
\label{eq:ks,mcccc}
  m_{max} \le m  \le m_{all}.
\end{equation}

Lastly, another approach that may be used for comparing and highlighting the efficiency in terms of number of integration points  of the proposed subspace-adaptive weights cubature rules  is the ``naive'' strategy of  applying independently the ECM (or an LP-based method) to each basis matrix. We will show in the ensuing discussion that the probability of this approach  in giving overlapping subsets of points  is rather low.

 \jahoHIDE{

\own{Furthermore, Fig. \ref{fig: concept gauss points} shows a visual example of a 1-D domain. In the figure, the 2 functions to integrate are shown to require 2 Gauss points each, as evidenced by the non-zero entries in the sparse vectors $\boldsymbol{\zeta}^{(i)}$. Moreover, the modified weights $\boldsymbol{\omega}^{(i)}$ contain 1 zero entry each, to account for the mismatch of the indices selected for the 2 different functions.}

\begin{figure}[H]
    \centering
    \includegraphics[width=.55\linewidth]{Notion_GaussPoints.pdf}
    \caption{ \revone{Visual example of a 1-D domain $\Omega = [0,1]$ discretized into 4 elements containing 2 Gauss points each, and 2 functions to integrate. The original weights vector $\boldsymbol{W}\in \mathbb{R}^{M}$ contains 8 entries, the same number of entries as the 2 sparse vectors $\boldsymbol{\zeta}^{(i)}$. In this example case, the selected unique Gauss point indices are $\Eind =\{1,2,7\}$, and the corresponding modified weights vectors $\boldsymbol{\omega}^{(i)}$ are shown to contain zero entries.}}
    \label{fig: concept gauss points}
\end{figure}
}

\revtwo{
\begin{remark}
The advantage of maximizing the number of integration points common to all subspaces becomes obvious in solid mechanics applications where the materials are governed by constitutive equations with internal variables—inasmuch as the number of integration points at which internal variables are to be tracked and stored is minimum. It is also advantageous in reduced-order implementations that can be characterized as ``noninvasive" in the sense that the transmission of information from one time step to the next is not done by updating the reduced coordinates, but rather through the nodal coordinates of the finite elements selected in the hyperreduction process. This is the case of the open-source software Kratos Multiphysics\footnote{\url{https://github.com/KratosMultiphysics/Kratos}}, employed in the hyperreduced-order models presented by the authors in Ref. \cite{bravo2024geometrically} and \cite{ares2023hyper}.
\end{remark}
}

\section{Numerical assessment}
\label{sec:Examples}

This section is devoted to the numerical assessment  of the proposed subspace-adaptive weights cubature/quadrature rules. Three examples will be used for this assessment.  The first two examples deal with the derivation of rules for univariate polynomial functions (actually the first example is the same employed for illustrating the notion of quadrature with subspace-adaptive weights  in Section \ref{sec:cubat.}).  The goal in both cases is to compare the performance (in terms of number of integration points), of the SAW-ECM introduced in Section \ref{sec: The Local Empirical Cubature Method}, and the  strategy based on Linear Programming, addressed in Section \ref{sec:LP}. In the latter case, several Linear Programming algorithms will be explored.   The (Python) files for reproducing these first two examples are publicly available on the GitHub repository: \url{https://github.com/Rbravo555/localECM}.

The third example is concerned with the application of the SAW-ECM algorithm to the local hyperreduction of a parameterized FE large-strains, structural model (for reasons that will become apparent later, the Linear Programming strategy will not be tested in this example).   Since the interest lies on evaluating the errors introduced exclusively by the cubature rule,  we will concoct a test in which other sources of errors are  either negligible or kept in check.

\subsection{  Scalar-valued polynomial functions }
\label{sec:examp1}

Let us consider the  set of functions $\{a^{(i)}\}_{i=1}^{k} : \Omega \times \mathcal{P} \rightarrow \mathbb{R}$ defined by:

\begin{equation}
\label{eq:7,ddddd}
    \begin{aligned}
       a^{(\mu+1)}: \ & [0,1] \times \mathbb{Z}_{+} \rightarrow \mathbb{R} \\
       \ & (x, \mu) \mapsto  x^{\mu} \ ,\\
    \end{aligned}
\end{equation}

\noindent where each function (monomial) is defined over the spatial domain $[0,1]$ with the set of non-negative integers $\mathbb{Z}_{+}$ as parameter space. The position of the candidate points in the spatial domain coincide with those of a 20-points Gauss rule. The parameters for this example are thus:
\begin{equation}
      N = 20 \hspace{5mm}  , \hspace{5mm} k=P=6 \hspace{5mm}  , \hspace{5mm} \{\bar{\boldsymbol{x}}_i\}_{i=1}^{20} \in [0,1] \hspace{5mm} ,  \hspace{5mm} \mathcal{P}^h = \{0,1,2,3,4,5\}  \ .
\end{equation}

\noindent The monomials and corresponding  basis functions\footnote{These basis functions are $L_2(\Omega)$ orthogonal, as discussed in Appendix \ref{sec:appendix 3}.} obtained from the SVD are shown in Figures \ref{fig: functions_ex1} and \ref{fig: basis_functions_ex1}, respectively.

\begin{figure}[!ht]
  \begin{minipage}[b]{0.5\linewidth}
    \centering
    \includegraphics[width=.92\linewidth]{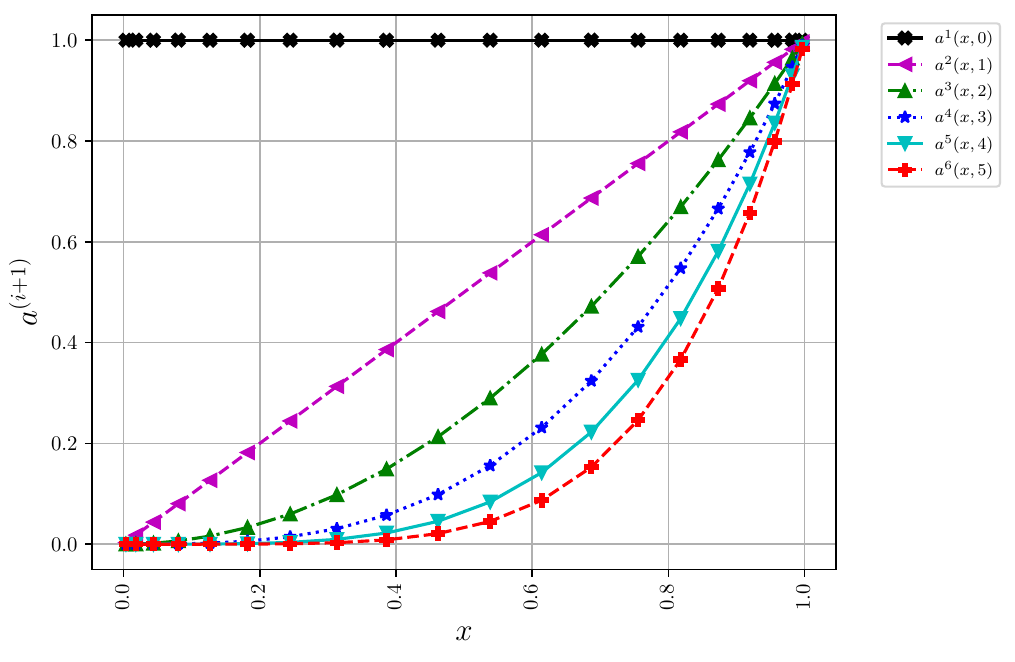}
    \caption{Functions $a^{(j)}: \Omega \times \mathcal{P} \rightarrow \mathbb{R}$}
    \label{fig: functions_ex1}
    \vspace{4ex}
  \end{minipage}
  \begin{minipage}[b]{0.5\linewidth}
    \centering
    \includegraphics[width=.91\linewidth]{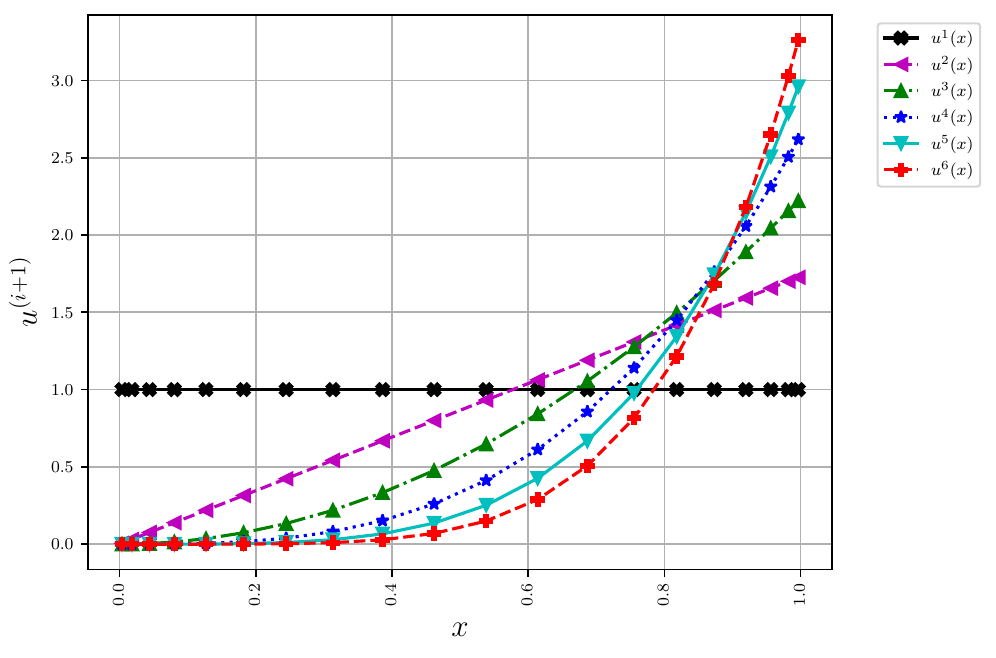}
    \caption{Basis functions $u^{(j)}: \Omega \rightarrow \mathbb{R}$}
    \label{fig: basis_functions_ex1}
    \vspace{4ex}
  \end{minipage}
\end{figure}

Application of the SAW-ECM   (Algorithm \ref{alg:ecm_kratos_ECM_local_pod}) results in just one point $(m=1)$ for the 6 subspaces; Figure \ref{fig: sparsity local ecm} shows the \emph{sparsity} plot (representation of the nonzero entries of a matrix  ) of the   vectors $\boldsymbol{\zeta}^{(i)}$ involved in the integration of each monomial (i.e. subspace). Thus, we see that  the SAW-ECM indeed provides an optimal solution to the problem (as discussed in Section \ref{sec:cubat.}, this problem possesses infinite optimal solutions ).
\begin{figure}[!ht]
    \centering
    \includegraphics[width=.45\linewidth]{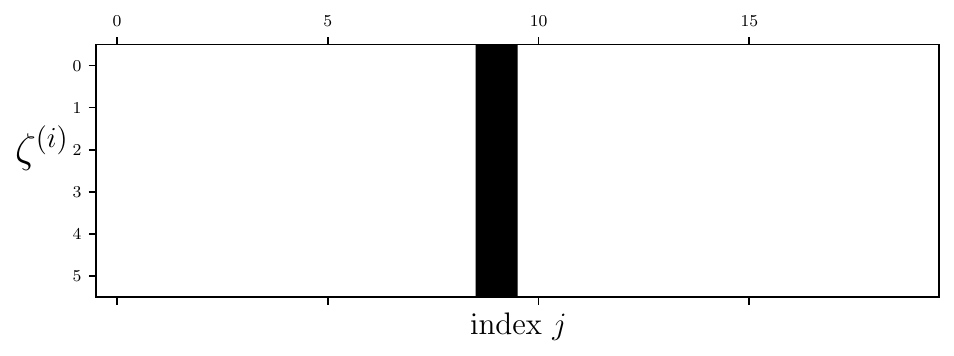}
    \caption{ Integration of the  6 subspaces of functions described by    Eq. \ref{eq:7,ddddd} (6   monomials).  Sparsity plot of the weight vectors  obtained by the SAW-ECM.   }
    \label{fig: sparsity local ecm}
\end{figure}
As for the Linear Programming-based strategy,  we construct the matrices and vectors $\tilde{\boldsymbol{U}}, \tilde{\boldsymbol{b}}, \tilde{\boldsymbol{\zeta}}$ as in   Eq. \ref{eq: lp vectors definition}, and then call the LP solver:
\begin{equation}
    \left(\Eind,   \{\boldsymbol{\omega}^{(i)}\}_{i=1}^6  \right)\leftarrow \texttt{LP} \left( \tilde{\boldsymbol{U}}, \mathbbm{1} , \tilde{\boldsymbol{b}},  \tilde{\boldsymbol{\zeta}} \succeq \boldsymbol{0} \right) \ .
\end{equation}
Optimization is performed using the \texttt{linprog} function provided by the \texttt{scipy.optimize} module in the SciPy library in Python. We employ the 6   methods available in this library; the solutions furnished by these methods  are shown in the sparsity plots of Fig. \ref{fig: sparsity linear programming}:
\begin{figure}[!ht]
    \centering
    \includegraphics[width=.8\linewidth]{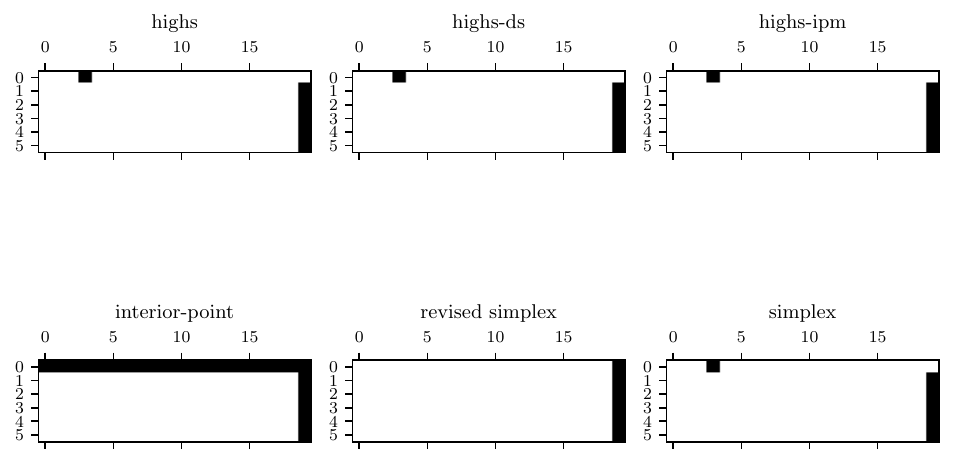}
    \caption{  Integration of the  6 subspaces of functions described by    Eq. \ref{eq:7,ddddd} (6   monomials). Sparsity plots furnished by the   {Linear Programming} strategy, using the 6   different methods available  when calling Python function   \texttt{linprog} ( \texttt{scipy.optimize} module).    }
    \label{fig: sparsity linear programming}
\end{figure}
It can be seen that, interestingly, the different tested LP algorithms furnish different solutions with different degrees of sparsity: from the interior point case, which leads to a dense solution using the 20 points, to the revised simplex, which delivers an  optimal sparse solution,  with just one point (notice that the position of this point is different from the one provided by the SAW-ECM in Figure \ref{fig: sparsity local ecm}).

To complete the assessment, we plot in   Figures  \ref{fig: sparsity global ecm} and  \ref{fig: sparsity independent ecm} the sparsity plots obtained by the ECM applied to the global subspace, and the ECM applied individually to each subspace, respectively. In the former case, we obtain $m_{all}=6$ points (as many as subspaces), and in the latter case, we get    one  single point per function, yet, as expected,   the indices of the selected points are not the same for all  the functions ---the  final number of selected point is $m= 4$.

\begin{figure}[!ht]
    \centering
    \includegraphics[width=.45\linewidth]{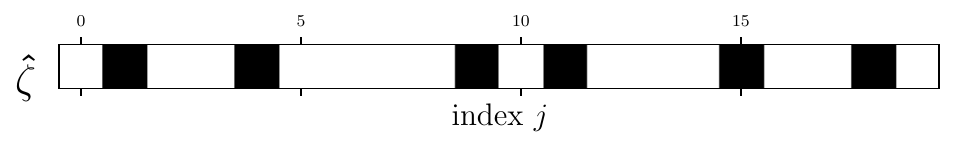}
    \caption{  Integration of the  6 subspaces of functions described by    Eq. \ref{eq:7,ddddd} (6   monomials). Sparsity plot corresponding to the application of the   ECM to the global subspace (i.e., same weights for the 6 subspaces).     }
    \label{fig: sparsity global ecm}
\end{figure}

\begin{figure}[!ht]
    \centering
    \includegraphics[width=.45\linewidth]{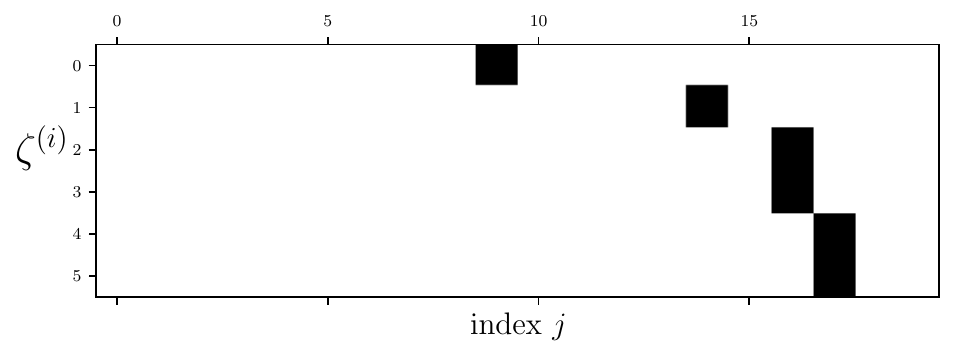}
    \caption{ Integration of the  6 subspaces of functions described by    Eq. \ref{eq:7,ddddd} (6   monomials). Sparsity plot for the case of applying,    independently,  the ECM to each  subspace. }
    \label{fig: sparsity independent ecm}
\end{figure}

 We summarize in Figure   \ref{fig: total number of points selected ex1} the result of the assessment in terms of number of points.  It may be concluded from this plot  that for this admittedly simplistic example, there is no apparent advantage in  adopting the Linear Programming strategy (using the revised simplex) or the SAW-ECM, for both give optimal solutions featuring one integration point.  However, the fact that the success of the  Linear Programming strategy is contingent upon the specific algorithm used  for solving the optimization casts doubts on  its reliability.   In the    studied case presented in the squel, in which we analyze  a vector-valued polynomial function, we  try to shed more light   on this issue.

\begin{figure}[!ht]
    \centering
    \includegraphics[width=.45\linewidth]{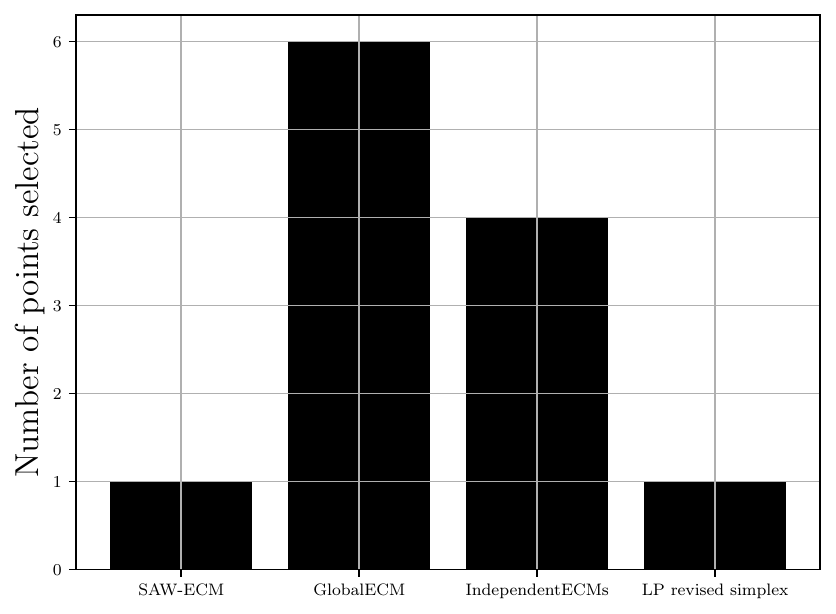}
    \caption{  Integration of the  6 subspaces of functions described by    Eq. \ref{eq:7,ddddd} (6 univariate monomials).  Total number of   points  employed  by each method. }
    \label{fig: total number of points selected ex1}
\end{figure}

\subsection{Vector-valued polynomial function}

Let us now consider the vector-valued polynomial function $\{a^{(i)}\}_{i=1}^{k} : \Omega \times \mathcal{P} \rightarrow \mathbb{R}^2$ defined by:
\begin{equation}
\label{eq:jmddddd}
    \begin{aligned}
       a^{(\mu+1)}: \ & [0,1] \times \mathbb{Z}_{+} \rightarrow \mathbb{R}^2 \\
        \ & (x, \mu) \mapsto  (1, x^{\mu}) \ .\\
    \end{aligned}
\end{equation}
The full-order integration rule is  a 50-points Gauss rule (one single finite element), and we take 20 samples of the parametric space; thus, the parameters defining this problem are:
\begin{equation}
      M = 50 \hspace{5mm}  , \hspace{5mm} k=P=20 \hspace{5mm}  , \hspace{5mm} \
    \{\bar{\boldsymbol{x}}_i\}_{i=1}^{50} \in [0,1] \hspace{5mm} ,  \hspace{5mm} \mathcal{P}^h = \{0,1,2,\dots,19\}  \ .
\end{equation}
The goal here is to determine the indices of the integration points $\Eind \subseteq \{1,2 \ldots 50 \} $ and the subspace-adaptive weights  $\omegaB^{(1)}, \omegaB^{(2)} \ldots \omegaB^{(20)}$ corresponding to the column spaces of the sample matrices $\A^{(1)}, \A^{(2)} \ldots \A^{(20)}$. The number of columns of each sample matrix is 2     (  $\A^{(i)} \in \RRn{50}{2}$), and except for $\A^{(1)}$, which has rank one, the remaining matrices \own{have}   rank 2. It follows then that,  according to the discussion in Section \ref{sec:ramk}, see inequality \ref{eq:ks,mcccc},     $m \ge 2$.

The sparsity plot corresponding to the solution provided by the SAW-ECM for the 20 sample matrices is shown in Figure \ref{fig: sparsity local ecm 2}. Observe that the set  $\Eind$ has only $m=2$ indices, which means that the SAW-ECM has achieved maximum efficiency. By contrast, the Linear Programming strategy, whose sparsity plots are displayed in Figure  \ref{fig: sparsity linear programming 2}, clearly fails to achieve this degree of efficiency: the solutions provided by the 6 LP methods    ranges  from the full dense solution obtained with the  interior-point method (50 points), to the 5-points solution furnished by the revised simplex.
\begin{figure}[!ht]
    \centering
    \includegraphics[width=.45\linewidth]{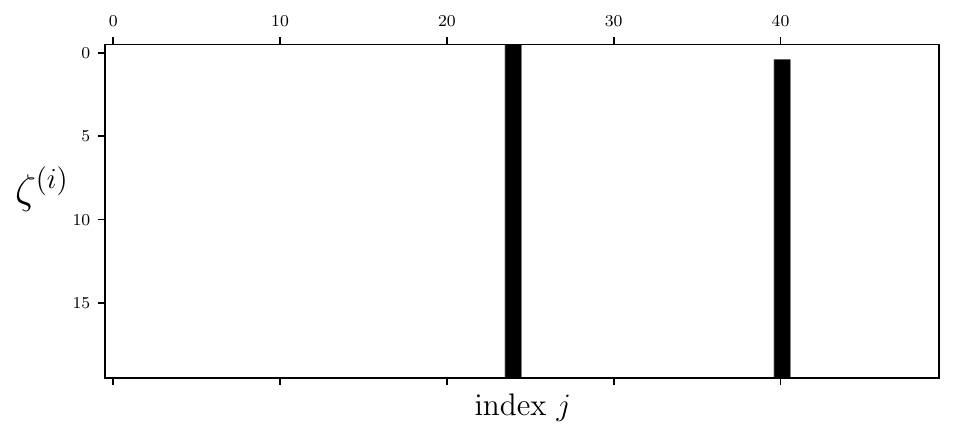}
    \caption{ Integration of the  20 subspaces of vector-valued polynomial functions described by    Eq. \ref{eq:jmddddd}.  Sparsity plot of the weight vectors  obtained by the SAW-ECM.  }
    \label{fig: sparsity local ecm 2}
\end{figure}
\begin{figure}[!ht]
    \centering
    \includegraphics[width=.8\linewidth]{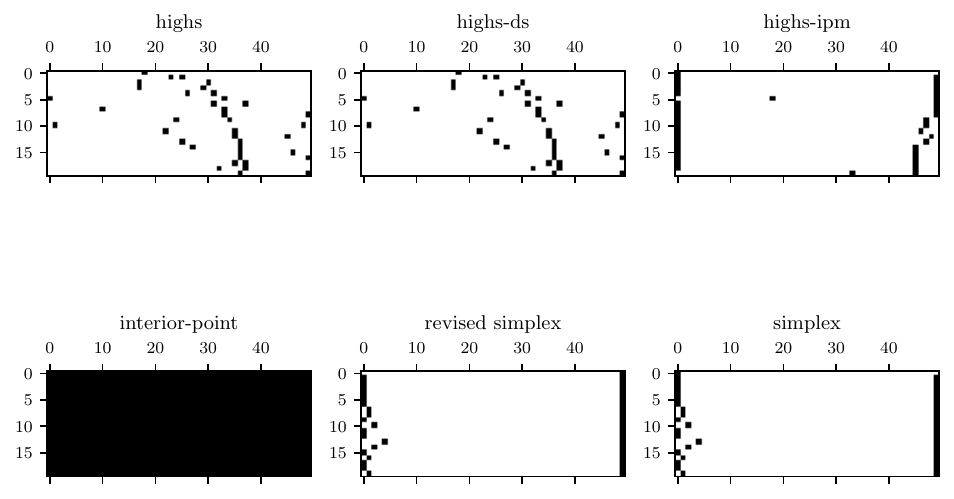}
    \caption{ Integration of the  20 subspaces of vector-valued polynomial functions described by    Eq. \ref{eq:jmddddd}.   Sparsity plots furnished by the   {Linear Programming} strategy, using the 6   different methods available  when calling Python function   \texttt{linprog} ( \texttt{scipy.optimize} module).}
    \label{fig: sparsity linear programming 2}
\end{figure}
Lastly, for completeness, we show in Figures \ref{fig: sparsity global ecm 2} and \ref{fig: sparsity independent ecm 2} the sparsity plots of the ECM applied to the global subspace and independently to each subspace, respectively. The former yields 20 points, and the latter 9 points.
 \begin{figure}[!ht]
    \centering
    \includegraphics[width=.45\linewidth]{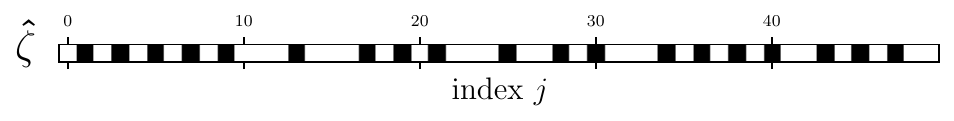}
    \caption{ Integration of the  20 subspaces of vector-valued polynomial functions described by    Eq. \ref{eq:jmddddd}. Sparsity plot corresponding to the application of the   ECM to the global subspace (i.e., same weights for the 20 subspaces). }
    \label{fig: sparsity global ecm 2}
\end{figure}
\begin{figure}[!ht]
    \centering
    \includegraphics[width=.45\linewidth]{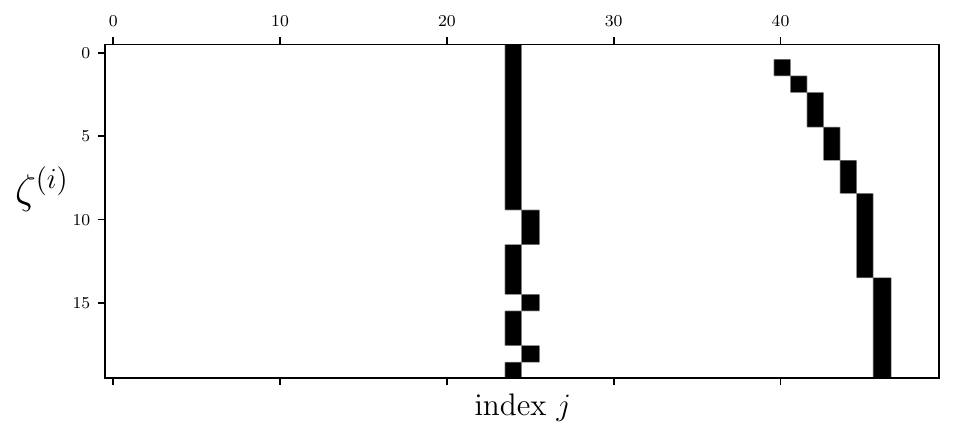}
    \caption{Integration of the  20 subspaces of vector-valued polynomial functions described by    Eq. \ref{eq:jmddddd}. Sparsity plot for the case of applying,    independently,  the ECM to each  subspace.  }
    \label{fig: sparsity independent ecm 2}
\end{figure}

\begin{figure}[!ht]
    \centering
    \includegraphics[width=.45\linewidth]{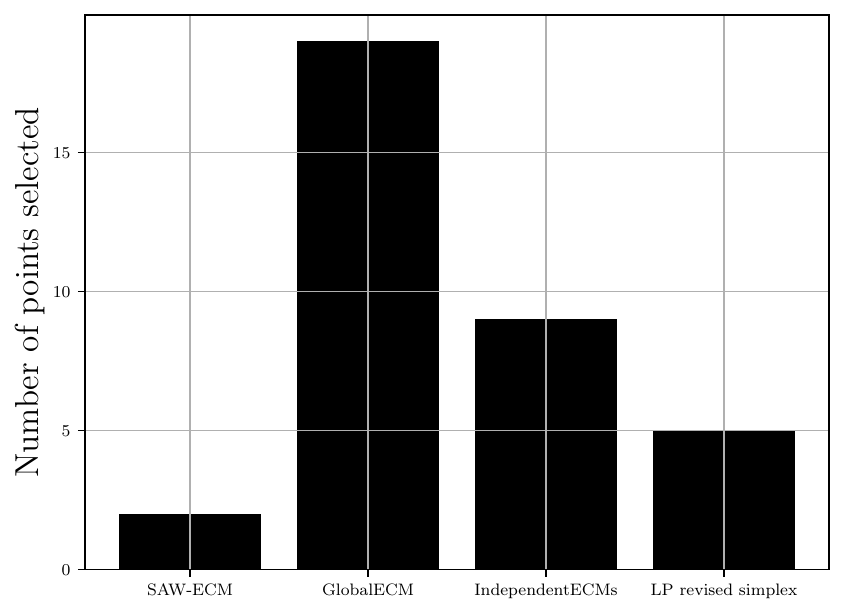}
    \caption{Integration of the  20 subspaces of vector-valued polynomial functions described by    Eq. \ref{eq:jmddddd}. Total number of   points  employed  by each method. }
    \label{fig: sparsity global ecm ex2}
\end{figure}

The summary of the results obtained for this family of functions  is presented in the bar plot of  Figure \ref{fig: sparsity global ecm ex2}.
These results confirm  the conjecture put forward in the previous section (the   Linear Programming strategy is not reliable because it gives varying number of points depending on the specific algorithm employed for solving the optimization problem\own{)}, and furthermore, provides evidence that, when dealing with more complex problems, the SAW-ECM is more efficient than the Linear Programming strategy (2 points versus 5 points, in the best case of the LP methods). For this reason, the last part of the assessment ---application to local hyperreduction of a finite element model, presented in the ensuing section--- will be carried out using solely the SAW-ECM.

The cause behind the higher efficiency of the SAW-ECM  over the Linear Programming strategy is not clear though. For the case of the global cubature problem, both methods   essentially    provide the same number of points when the sample matrix is  full rank ---this is empirically demonstrated by the authors in the introduction of  Ref. \cite{hernandez2024cecm}.  Thus, it is not apparent why this   is not inherited when moving to  the case of multiple subspaces, and, furthermore why the differences in performance vary depending on the particular LP algorithm employed for solving the optimization  problem. Further research is necessary to clarify this issue.  One plausible explanation might be  related  to the fact that  the  solution strategy behind the SAW-ECM exploits the requirement that the full-order weights  $\W$ must be strictly positive --- the notion of conical hull introduced in Section \ref{sec: non-uniqueness of ECM} relies on this very hypothesis ---whereas the LP strategy does not leverage this property at all in any of its steps.

\subsection{Local hyperreduction  of a parameterized finite element model}
\label{sec:localH}

\subsubsection{Problem setup}

The problem  chosen for illustrating the use of subspace-adaptive weights cubature rules (more specifically, the SAW-ECM, see Algorithm \ref{alg:ecm_kratos_ECM_local_pod}) in the context of  local hyperreduction   of  parameterized finite element analyses  is  a plane-strain  equilibrium problem of a $2 \times 2$ m      structure  with a centered circular void of diameter 1 m (see Figure \ref{fig:FIG_MESH}.a). The boundary conditions are    prescribed displacements on its external boundary; these displacements    are   parameterized in terms of 3   variables, namely, $E_{x}^M$ and $E_y^M$, which characterize  elongations along the \own{$x$}-axis and \own{$y$}-axis, respectively,   and $E_{xy}^M$, which describes the shear deformation of the structure.  The material obeys a Neo-Hookean constitutive law,  with  Young's Modulus $E = 70000$ MPa and a Poisson's ratio $\nu = 0.29$.  The finite element mesh, also displayed in  Figure \ref{fig:FIG_MESH}.a,  comprises 1005 bilinear quadrilateral elements, with 4 Gauss point per element (thus the total number of Gauss points is $M = 1005 \cdot 4 =  4020$); on the other hand,  the number of nodes is 1116, out of which $n_{unc}$ = 992 are interior nodes.  The goal is, given   $\boldsymbol{\mu} = \E^C =[E_x^C,E_y^C,E_{xy}^C]$, solve the        equilibrium  problem (in the large strain regime), and determine the  nodal displacements  $\d$.  Since there are no external forces, the  equilibrium problem for the case of the full-order (finite element) model   boils down to equating to zero the vector of  nodal internal forces $\F_{int} \in \Rn{N}$:
\begin{equation}
\label{eq:9ds..dd}
 \Fint(\d; \E^C)= \int_{\Omega_0}   \boldsymbol{B_0}^T  \boldsymbol{P}(\d;\E^C)  dV = \boldsymbol{0}.
\end{equation}
Here $\boldsymbol{P}: \Omega_0 \rightarrow \Rn{4}$ stands for the 1st Piola-Kirchhoff stress vector, whereas $\boldsymbol{B_0}: \Omega_0 \rightarrow \RRn{4}{N}$ is the strain-displacement matrix, in its globally supported format, relating the deformation gradient  at a given point in the undeformed domain $\Omega_0$ with  the  nodal displacements   ($\d \in \Rn{N}$)   corresponding to the unconstrained DOFs (which in this case are all the interior DOFs;   $N = n_{unc} \cdot 2 = 992 \cdot 2 = 1984$).

It is worth noting that the above described  problem is akin to    the  classical   homogenization problem  in large strains, which is  one of the   multi-query scenarios in which    the use of  reduced-order modeling is warranted, see for instance Refs. \cite{yvonnet2007reduced, hernandez2014high, oliver2017reduced, caicedo2019high}. The input parameters $\E^C$ would play the role of macroscopic Green-Lagrange strain vector, and the output of interest would be the volume average of the 1st Piola-Kirchoff  stresses ($\P^C = 1/V \int_{\Omega_0} \P d \Omega )$.

\subsubsection{Statement of the local hyperreduction problem }
\label{sec:state}

Let us now address the problem of constructing a local hyperreduced-order model for the parameterized FE problem represented  by Eq.  \ref{eq:9ds..dd} ---a sketch of this process was already given  in Section \ref{sec:j,lllodddd} when motivating the need for subspace-tailored weights cubature rules.   Following the standard offline/online computation paradigm, we solve in the offline stage equation \refpar{eq:9ds..dd}  for a given set  of training parameters, resulting in $P$ solution snapshots $\d^1, \d^2 \ldots \d^P$.    These $P$ snapshots are grouped according to some criterion into $k$    clusters, with $P_i$ snapshots each \own{(}$1 \le k \le P$ is a user-prescribed parameter\own{)}, and then the truncated SVD is applied to each cluster separately, giving local basis matrices $\PhiB^1, \PhiB^2 \ldots \PhiB^k$  ($\PhiB^i \in \RRn{N}{n_i}$, $n_i \le P_i$). The state of the system (nodal displacements $\d$)   is described  by indicating in which cluster  resides  the solution, and the amplitude (or reduce coordinates) $\q^i \in \Rn{n_i}$ of the basis matrix corresponding to this cluster (that is, $\d = \PhiB^i \q^i$). Once we know the cluster, and  given an input parameter $\E^C$, the reduced coordinates $\q^i$ are determined by solving the nonlinear equation resulting from projecting   \refeq{eq:9ds..dd}  onto  the span of $\PhiB^i$, that is, ${\PhiB^i}^T \Fint(\q^i;\E^C) = \zero$. This is when the subspace-tailored weight rule proposed in this work  comes into play. The integral corresponding to this projected equation will be approximated by
\begin{equation}
\label{eq:9dsd..d*d}
  {\PhiB^i}^T \Fint =   \int_{\Omega_0}   (\boldsymbol{B_0}(\x) \PhiB^i)^T  \P(\x,\PhiB^i \q^i)  dV \approx \sum_{g=1}^m  \omega^i_g (\boldsymbol{B_0}(\x_g) \PhiB^i)^T   \P(\x_g,\PhiB^i \q^i).
\end{equation}
Here  $\x_g = \xBAR_{\Eind{g}}$  ($\xBAR_j$ is the $j$-th Gauss point of the mesh, $j=1,2 \ldots M$)   and $\{\omega^i_1,\omega^i_2, \ldots \omega_m^i \}$ ($i=1,2 \ldots k$) are the sets of nonnegative weights adapted to each particular subspace. Both the reduced set of  indices $\Eind \subseteq \{1,2 \ldots M\}$    and the adaptive weights are determined by using the SAW-ECM described in Algorithm \ref{alg:ecm_kratos_ECM_local_pod}. The inputs of this algorithm are the vector of Gauss weights  $\W \in \Rn{M}$ \own{(}recall that these weights also include the Jacobian of the mapping from parent configuration to physical configuration\own{)}, and the integrand matrix  for each cluster:  $\A^1, \A^2 \ldots \A^k$. The integrand matrix of the $i$-th cluster is computed, in the offline stage as well, by the procedure described in the following.

Let $\d^{b_1}, \d^{b_2} \ldots \d^{b_{P_i}}$ be the $P_i$ displacement snapshots used for constructing the basis matrix $\PhiB^i$ (here $\{b_1,b_2 \ldots b_{P_i} \} \subseteq \{1,2 \ldots P \} $\own{)}.  For each of these snapshots, we compute its projection onto the basis matrix, i.e., $\dTILDE^{ \,b_j} = \PhiB^i {\PhiB^i}^T \d^{b_j}$  ($j=1,2 \ldots P_i$);    then, for each  Gauss point $\xBAR_g$ of the mesh ($g=1,2 \ldots M$),  we evaluate the corresponding stresses through the constitutive equation,  $\P = \P(\xBAR_g,\dTILDE^{b_j})$, and multiply these stresses by the transpose of the matrix of virtual deformation at the same Gauss point ($\boldsymbol{B_0}(\xBAR_g) \PhiB^i$). In practice,   the strain-displacement  matrix in a FE implementation  is given at element level, so this matrix is actually calculated as  $\B_0^{(e)}(\xBAR_g) {\PhiB^{(e)}}^i$, where $\B_0^{(e)}$ denotes the strain-displacement matrix of the element $\Omega_0^e$ containing point $\xBAR_g$, and ${\PhiB^{(e)}}^i$ is the restriction of $\PhiB^i$ at the DOFs of such  element $\Omega_0^e$.  The resulting integrand matrix $\A^i$ (whose entries  represent physically internal virtual work per unit volume) will be thus of size $M \times P_i  n_i$.

\begin{figure}[!ht]
    \centering
    \includegraphics[width=0.99\linewidth]{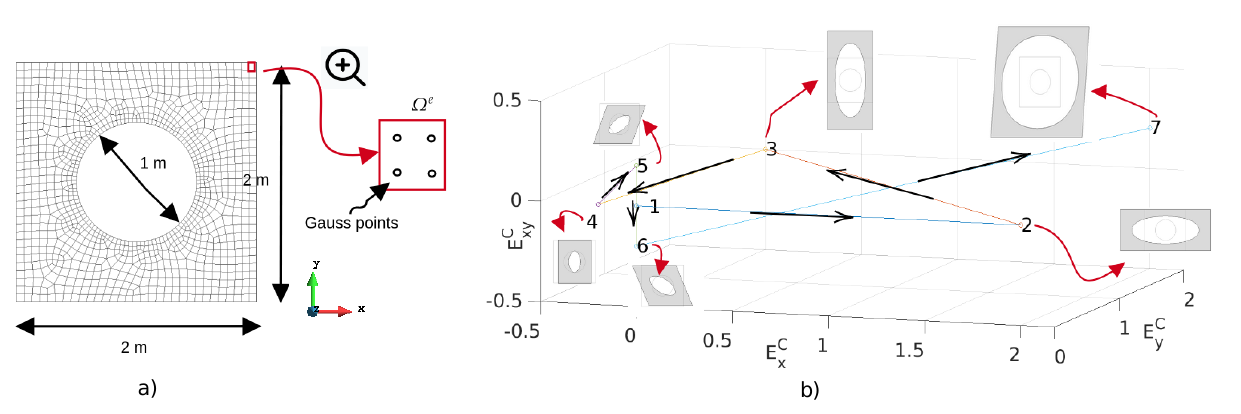}
    \caption{a) Finite element mesh of the structure under study, with $1005$ bilinear quadratic elements, and 4 Gauss points per element (a total of $M = 4020$  Gauss points).  b) Piecewise affine trajectory in parameter space used to train the HROMs, along with the deformed shapes of the structure at the   points defining each segment (  the reference configuration is   shown as well in each case, in order to appreciate the extent of the deformations).  The axes of the graph represent a parametrization  of the displacements of the external boundary   $\E^C = [E^C_x,E^C_y,E^C_{xy}]$, where   $E_x^C$ is the elongation along the $x$ axis, $E_y^C$ is the elongation along the $y$ axis, and $E_{xy}^C$ represents shear deformations. The points defining the trajectory are $\E^C_1 = [0,0,0]$, $\E^C_2 =[2,0,0]$, $\E^C_3 =[0,2,0]$, $\E^C_4 =[-0.2,0,0]$, $\E^C_5 =[0,0,0.2]$, $\E^C_6 =[0,0,-0.2]$ and $\E^C_7 =[2,2,0.2]$.    }
    \label{fig:FIG_MESH}
\end{figure}

\subsubsection{Transition errors and scope of the assessment}
\label{sec:trans}

In describing     the     problem represented by Eq. \refpar{eq:9dsd..d*d}, we have glossed over one    fundamental, challenging  question which pervades local HROMs: how to group  the   snapshots in clusters  so that  the transition, in the online stage,  between neighboring clusters  is made without incurring in additional errors.  Suppose that, at a certain time of the online analysis, the solution vector $\d^{old}$ moves from a cluster with basis matrix $\PhiB^{old}$  (i.e. $\d^{old} = \PhiB^{old} \q^{old}$\own{)} to a cluster with basis matrix $\PhiB^{new}$.     Since the basis matrices are by construction  orthogonal (${\PhiB^i}^T \PhiB^i = \ident$), the best     representation ---in the least-squares sense---that one can obtain in the new cluster is given by $\d^{new} =  \PhiB^{new} (\PhiB^{new^T} \PhiB^{old}) \q^{old} $  \own{(}$\d^{new}$ is thus the orthogonal projection of $\d^{old}$ onto the span of $\PhiB^{new}$). It follows from this expression  that error-free transitions requires $\PhiB^{old} \q^{old} \in \spanb{\PhiB^{old}} \cap \spanb{\PhiB^{new}}$, i.e., the ``old'' state should pertain to the intersection of the spans of the basis matrices. Failure to meet this condition will unavoidable result in transitions errors, the worst conceivable scenario  being when both subspaces are orthogonal ($\PhiB^{new^T} \PhiB^{old} = \zero$), since in this scenario  all the information encoded by the reduced coordinates $\q^{old}$ is lost upon changing the cluster. Thus, in the offline stage, when grouping the snapshots into clusters,  care is to be exercised so that all clusters that may be considered neighbors in some metric have nonzero intersections. Researchers in the field have attempted to enforce this condition  by introducing \emph{cluster overlapping},  which implies that some snapshots may pertain simultaneously to different clusters, see e.g. Refs.  \cite{pagani2018numerical,grimberg2021mesh,washabaugh2012nonlinear,amsallem2012nonlinear}. The problem is elusive though, because overlapping criteria, no matter how sophisticated, can only control  transition errors when  the   input parameters are the same   used  to train  the local HROM.   In general, for testing parameters different from the training parameters, there is certain degree of  uncertainty in this regard, and transition errors will be present to a  greater or lesser extent depending on the level of overlapping ---of course, one can always increase the  overlapping, but this will increase the size of the clusters, and   the benefits of using local HROMs would be eventually lost.

 Studying the impact on such transition errors in the hyperreduction process, although of utmost importance for building local HROMs with online predictive capabilities, is out of the scope of the present work. Accordingly, in what follows, we
  focus
only on the consistency of the local HROMs (consistency, as defined in Ref. \cite{carlberg2011efficient}, is  the capability to  predict with any desired degree of accuracy the same data used for training the model\own{)}.  Objectionable as this assessment  might be from the machine learning perspective, it is the only way of ensuring that transition errors are negligible, and therefore, of examining the errors   that are  caused exclusively by  the approximation of the spatial integral in \refeq{eq:9dsd..d*d} via the proposed cubature method.

    We train the local HROMs using a single, piecewise affine trajectory in parameter space,   displayed in   Figure \ref{fig:FIG_MESH}.b. \revone{For carrying out  the incremental, nonlinear analysis, this trajectory is discretized into 3497 steps (total Lagrangian formulation).    At each step, the   system of nonlinear equilibrium equations Eq.(\ref{eq:9ds..dd})  is solved   using a Newton-Rahpshon iterative scheme with convergence tolerance $\epsilon_{nr} = 10^{-10}$. }  On the other hand, } the clustering of the corresponding $P= 3497$ displacement snapshots   into distinct groups  ---so that no   transition errors are introduced---  proves trivial in this problem,  for  it suffices to take  overlapping sequences of consecutive snapshots, with an overlap of    one snapshot with the next and previous clusters in the sequence.

 \subsubsection{Case with just one subspace (global HROM) }
 \label{eq:9.222222}

\jahoHIDE{See Matlab file \url{/home/joaquin/Desktop/CURRENT_TASKS/MATLAB_CODES/TESTING_PROBLEMS_FEHROM/CLUSTERING_HROM/PAPER_SAW_ECM.mlx}}

\begin{figure}[!ht]
    \centering
    \includegraphics[width=0.99\linewidth]{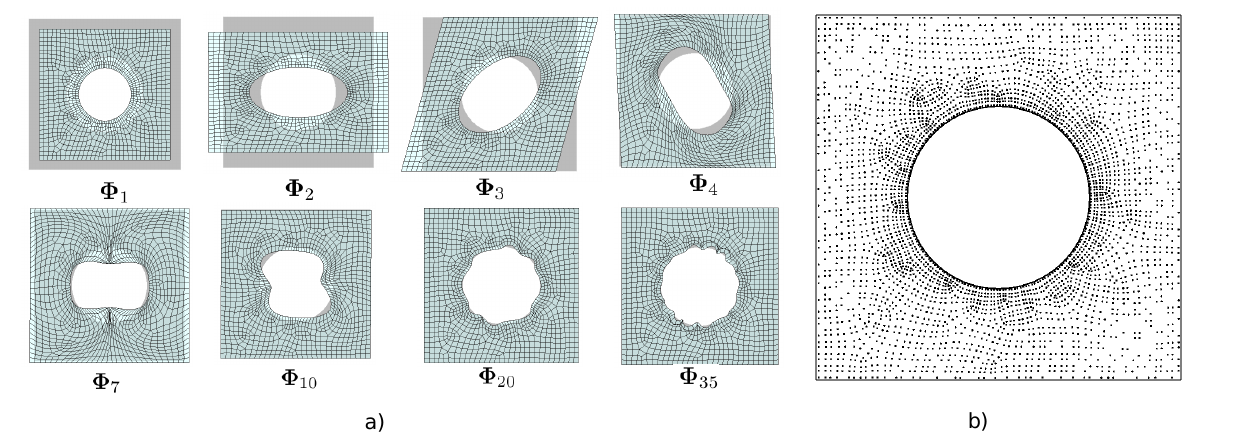}
    \caption{Case with just $k=1$ cluster (global HROM). a) Deformed shapes corresponding to 8   vectors out of the $n=50$ basis vectors obtained via the SVD.  b) Position (in the undeformed configuration) of the $m = 3443$ Gauss points (out of $M= 4020$   points)  selected by the   Empirical Cubature Method (Algorithm \ref{alg:ecm_global}).    }
    \label{fig:FIG_MODES}
\end{figure}

     Let us begin by studying the limiting  case of using just one cluster ($k=1$) ---notice that this case corresponds to  the standard,  global HROM, because there is only one subspace of functions to be integrated.  The tolerance for the truncated SVD for determining the basis matrix $\PhiB^{glo}$ is set to $\epsilon = 10^{-8}$. This SVD yields $n = 50$ left singular vectors; for illustrational purposes, we plot  the deformed shapes corresponding to some of these basis modes   in Figure \ref{fig:FIG_MODES}.        Following the procedure described previously in Section  \ref{sec:state}, we compute the integrand matrix $\A$, which has in this case $M = 4020$ rows and $n \cdot P = 50 \cdot 3497 = 174850$ columns. Because of its relative large size,  the SVD of this matrix (see expression \ref{eq:svd of A})   is carried out by the \emph{Sequential Randomized SVD} proposed by the authors in Ref. \cite{hernandez2024cecm}, appendix\footnote{ A matlab version of this algorithm can be found in the repository \url{https://github.com/Rbravo555/CECM-continuous-empirical-cubature-method/blob/main/CODE_CECM/SVDlibrary/SRSVD.m}} A, with an   error tolerance    $\epsilon_{svd} = 0$ ---this zero error tolerance   guarantees that the cubature problem is solved exactly, and therefore, the only   error introduced in the local HROM is to be attributed to the truncation error in  the first stage of reduction.  This SVD reveals that there is a total of 3442  left singular vectors. However, problem \refeq{eq:fffff} would prove ill-posed if we simply form the basis matrix $\U$ using this 3442 left singular vectors, because under such circumstances  $\d = \Fint =  \zero$  by virtue of \refeq{eq:9ds..dd} ---and, consequently, any set of indices $\Eind$ with weights $\omegaB = \zero$ would be a solution of problem \refeq{eq:fffff}.  As we mentioned in passing in Section \ref{sec:LCO}, and reflected in Line \ref{line:augment} of Algorithm \ref{alg:ecm_kratos_ECM_local_pod}, the remedy to this ill-posedness  is to add one column to the matrix of left singular vectors so that the space of constant functions is exactly integrated ---while keeping the orthogonality of the integrand modes. If orthogonality is in the Euclidean sense, this amounts to simply adding a column of ones ---the reader is referred to Appendix \ref{sec:appendix 2} to see how to proceed in a general case.  In doing so,   $\d= \U^T \W =  [\zero^T,V]^T$, $V = \sum_{g=1}^M W_g$ being the volume of the domain, and the problem ceases to be ill-posed. Using as input the resulting matrix $\U$ of 3443 columns,   the ECM of Algorithm \refpar{sec:LCO}   gives exactly $m= 3443$   integration points (the location of these points are shown in Figure \ref{fig:FIG_MODES}.b).

     The accuracy in reproducing with the resulting HROM (which has $n=50$ DOFs and $m= 3443$ integration points) the   trajectory used for training is measured by computing the Frobenius norm of the difference between the   matrix of original FE displacement  snapshots and its  HROM counterparts (divided by the Frobenious norm of the former). This yields an error of $e_{glob} = 1.1 \cdot 10^{-8}$; we see thus that the a posteriori error is     only 1.1 times larger than the a priori error prescribed by the truncation tolerance of the SVD for displacements.   Therefore, this HROM can be deemed consistent in the sense given in Ref. \cite{carlberg2011efficient}. However, it is clear that the
       model does not deserve the  qualifier ``hyperreduced'', because the second reduction in terms of integration points is marginal: $M/m = 4016/3443 = 1.17$.        The reason for this poor hyperreduction is    that, firstly, the number of basis vectors for displacements is relatively high ($n = 50$), and secondly,
     the dimension of the subspace where the integrand  (internal work per unit volume) lives scales, at least,  with   $n^2$.    This follows from the fact that the integrand function in \refeq{eq:9dsd..d*d} is the product of the $n$   strain modes $\B_0 \PhiB^{glo}$  and the 1st Piola-Kirchhoff stresses $\P$ engendered by this very $n$ strain modes. The stresses thus live in a space of dimension $n^{stress} = \ORDER{n^{1+\alpha}}$, and consequently, the integrand itself resides in a space of dimension $\ORDER{n^{2+\alpha}}$, where $\alpha \ge 0$ depends on the degree of nonlinearity (which in this case is relatively high, for the size of the final deformed structure is 200 \% larger than its undeformed configuration, see Figure \ref{fig:FIG_MESH}.b).

      \subsubsection{Case with maximum number of subspaces }
\label{sec:caseMAX}

\begin{figure}[!ht]
  \centering
  \subfigure[]{\label{fig:modes}\includegraphics[width=0.45\textwidth]{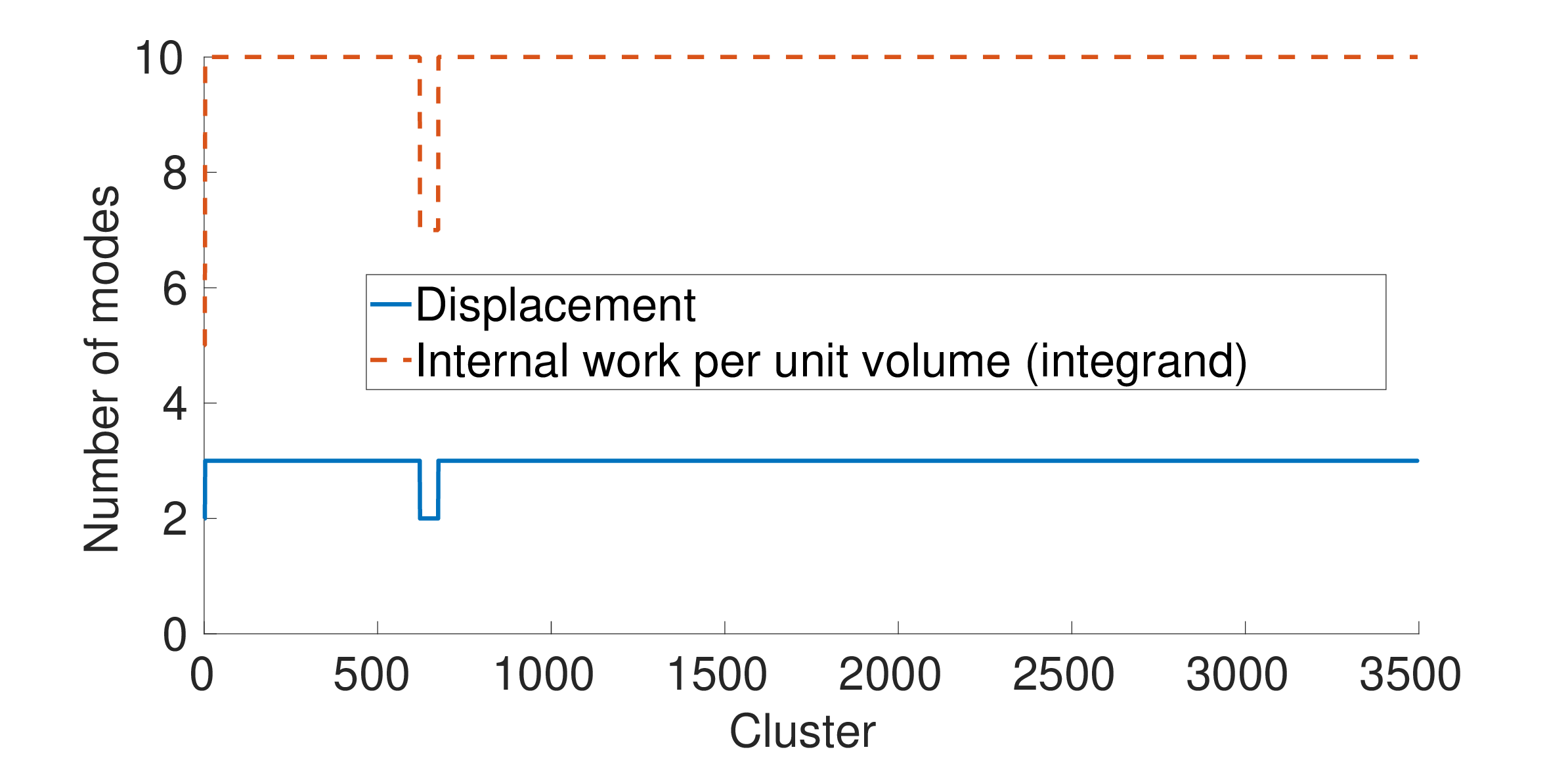}}
   \subfigure[ ]{\label{fig:calls}\includegraphics[width=0.45\textwidth]{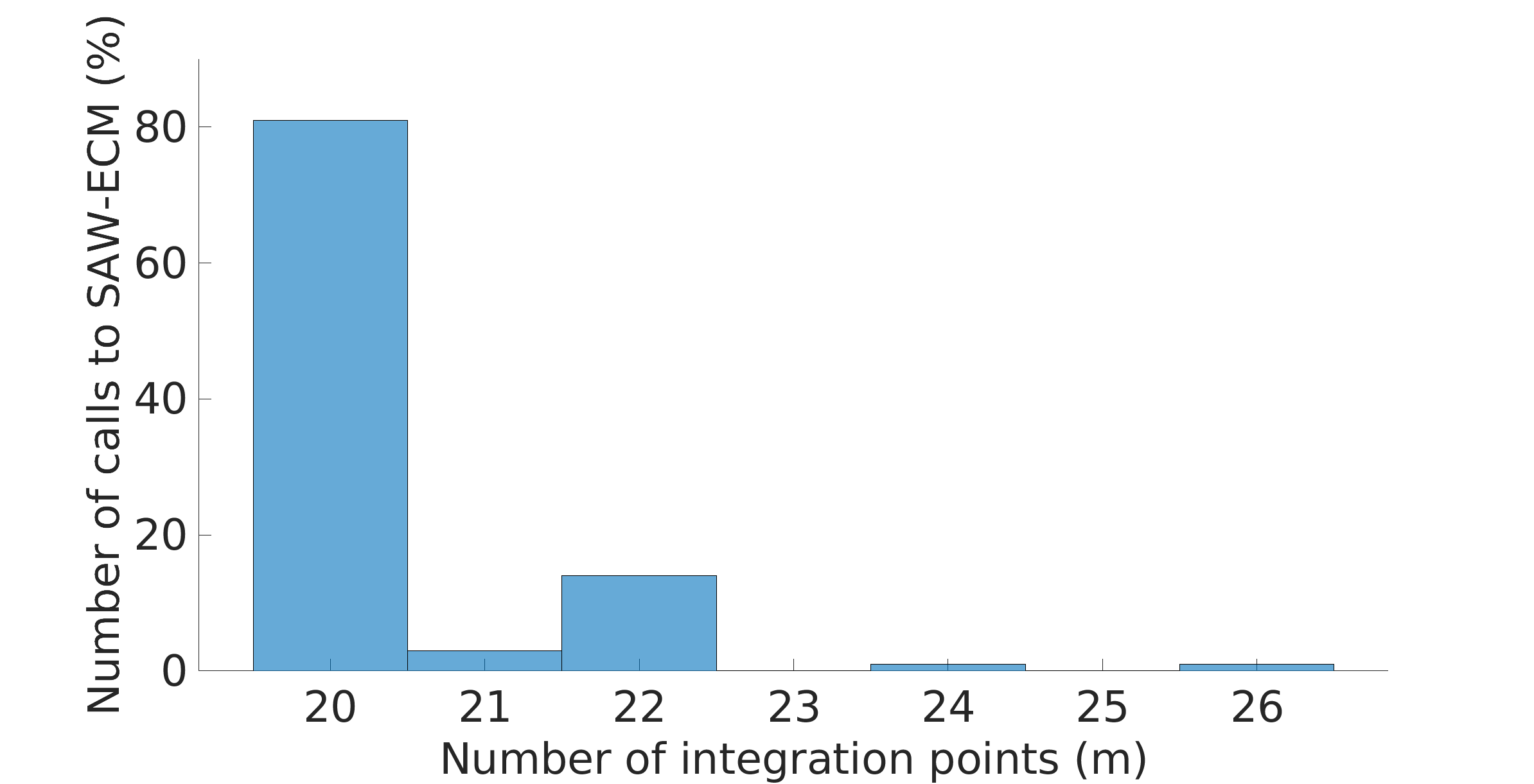}}
      \subfigure[ ]{\label{fig:iter}\includegraphics[width=0.45\textwidth]{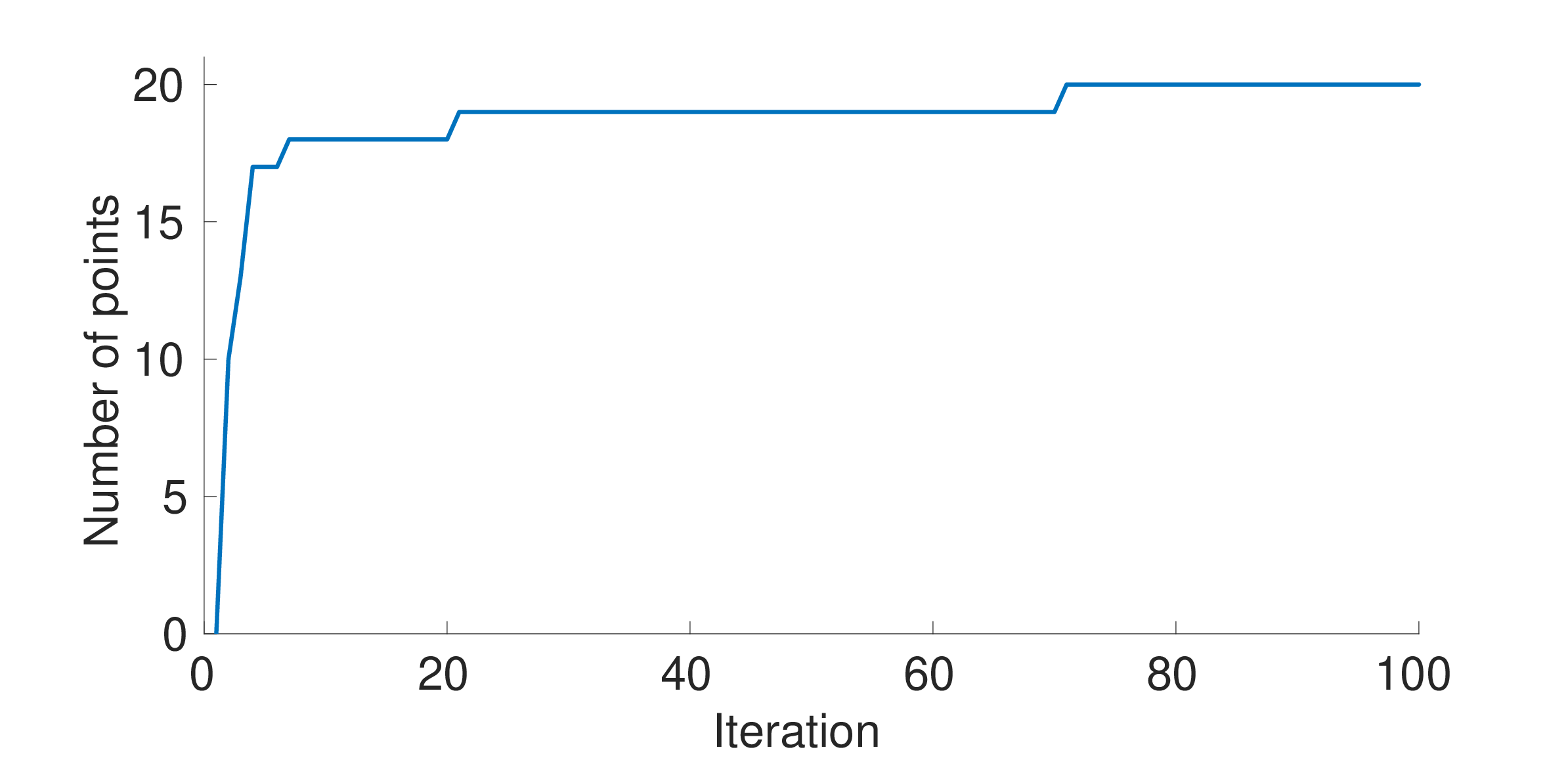}}
            \subfigure[ ]{\label{fig:loc}\includegraphics[width=0.45\textwidth]{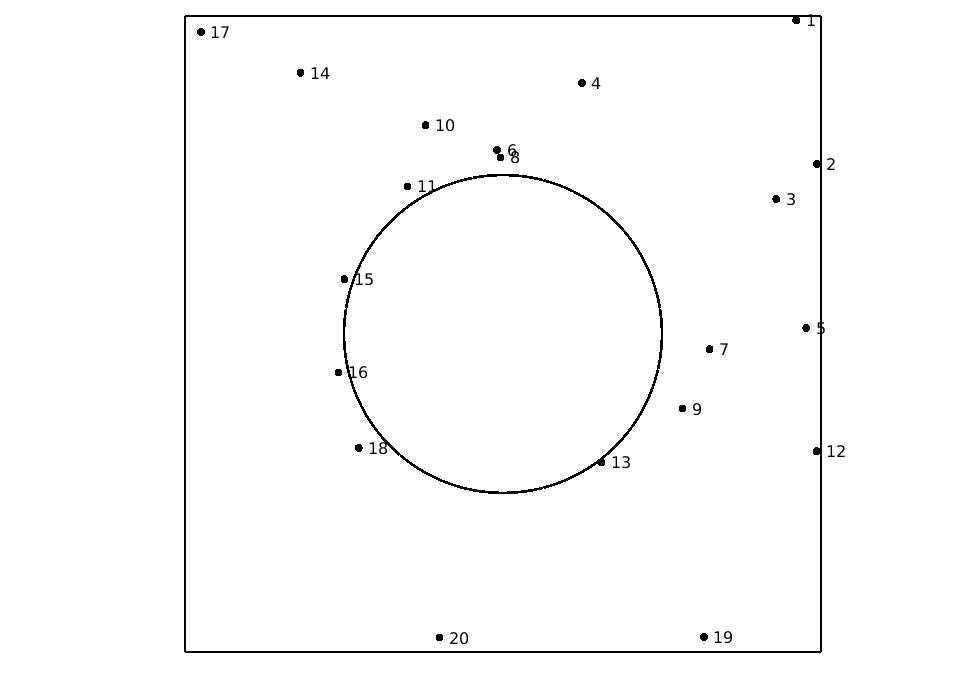}}
   \caption{ Case with $k=3495$ clusters (each cluster is  formed by 3 consecutive snapshots).  a) Number of  modes for displacements    and for  the integrand function.  b) The number of common points for all   clusters (i.e., subspaces) delivered by the SAW-ECM depends on the order in which the basis matrices in Algorithm \ref{alg:ecm_kratos_ECM_local_pod} are processed (see Line \ref{line:saw5}).  This bar plot depicts the probability distribution obtained from  100 random permutations of $\{1,2 \ldots 3495 \}$.  c) Number of iterations in the loop over clusters (Line \ref{line:saw1} in Alg. \ref{alg:ecm_kratos_ECM_local_pod}) required to achieve the final candidate set  (for one of   the cases with $m=20$ points).  d) Location of the $m=20$  points selected in  such a case.  The value of the adaptive weights at such points for  different clusters can be seen in Figures \ref{fig:FIG_WEIGHTS} and \ref{fig:FIG_TRANSW}.   }
  \label{fig:FIG_8}
\end{figure}

     Having discussed the (poor) levels of hyperreduction achieved when using the global HROM (just one cluster), let us now move to the other extreme in the clustering spectrum, which is when  we use almost as many clusters as snapshots.  As pointed out previously, zero-error transitions demand taking    sequences of   consecutive snapshots with an overlap of    one snapshot with the next and previous clusters in the sequence.  It follows then  that, in  this limiting case,  the clusters will be   formed by groups of 3 snapshots:  $\{\d^1,\d^2,\d^3 \}$, $\{\d^{2},\d^{3},\d^4\}$   $\ldots \{\d^{i-1},\d^{i},\d^{i+1}\}$, $\ldots \{\d^{P-2},\d^{P-1},\d^{P} \}$. The total number of clusters (and  thus of  subspaces of functions to be integrated) is equal to $k = P - 2 = 3495 $.      Application of the SVD  (with the same truncation tolerance as in the global case, $\epsilon = 10^{-8}$) to each group of 3 snapshots produces basis matrices, see Figure \ref{fig:modes}, with   3 modes    except for the 2 first clusters, on the one hand,  and  clusters from 622 and 671, on the other  hand, which have 2 modes (clusters from 622 and 671 corresponds to the last portion of the trajectory going from point 1 to point 2 in Figure \ref{fig:FIG_MESH}.b). With   $\PhiB^1,\PhiB^2 \ldots \PhiB^k$ at our disposal, we compute next,  by the procedure outlined in Section \ref{sec:state}, the integrand matrices $\A^1, \A^2 \ldots \A^{k}$ associated to these basis matrices, then use the SVD (with $\epsilon_{svd} = 0$) to determine the left singular matrices of each integrand matrix, and finally augment each of these left singular matrices with an additional column so that each local cubature problem is well-posed. This results in the distribution of number of  integrand modes per cluster shown in Figure \ref{fig:modes}: except for the first two clusters, which have $m_1=m_2 = 5$ modes, and clusters from 622 and 671, which have 7 modes, the remaining clusters have $3^2 + 1 = 10$ modes.    Thus, we can assert,   in the light of the discussion presented in Section \ref{sec:examp1}, that the   number of common points for all clusters  that can be obtained by the SAW-ECM is bounded below by $m_{max} = 10$.  It is worth examining as well what is the upper bound  $m_{all}$ for this case, which, recall, is the dimension of the global subspace, formed as the sum of the spans of all the basis matrices.   To this end,  we concatenate  all basis matrices in a single matrix: $\U = [\U^1,\U^2 \ldots \U^P]$, and apply the untruncated SVD to determine its rank.   This yields  $m_{all} = M = 4020$ ---the matrix has thus full rank. This implies   that the standard  hyperreduction  approach  of using the same set of weights    for  all clusters (see discussion in Section \ref{sec:origin}) offers no benefit in this case in terms of reduction of computational cost  ---the number of integration points is  equal to the number of Gauss points of the mesh, and hence there is no hyperreduction at all.

            \begin{figure}[!ht]
  \centering
  \subfigure[Cluster $1$  ($\E^C_1 = (0,0,0)$)]{\label{fig:cl1}\includegraphics[width=0.49\textwidth]{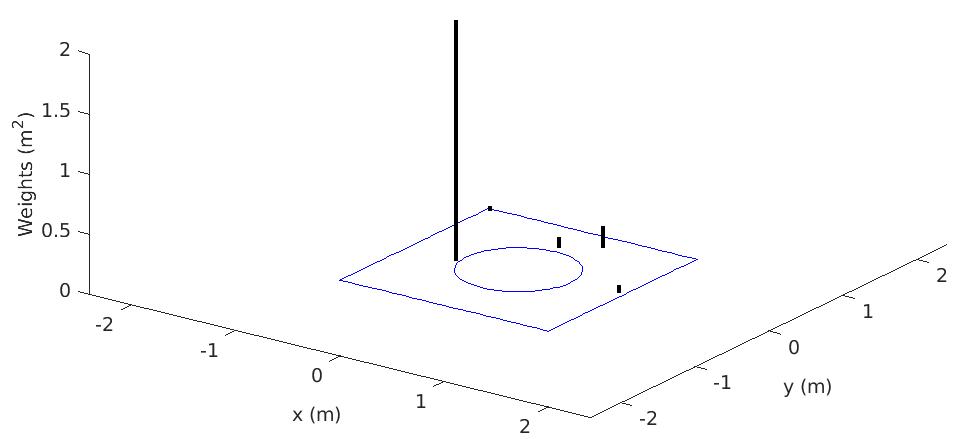}}
         \subfigure[Cluster $675$  ($\E^C_2 = (2,0,0)$) ]{\label{fig:cl2}\includegraphics[width=0.49\textwidth]{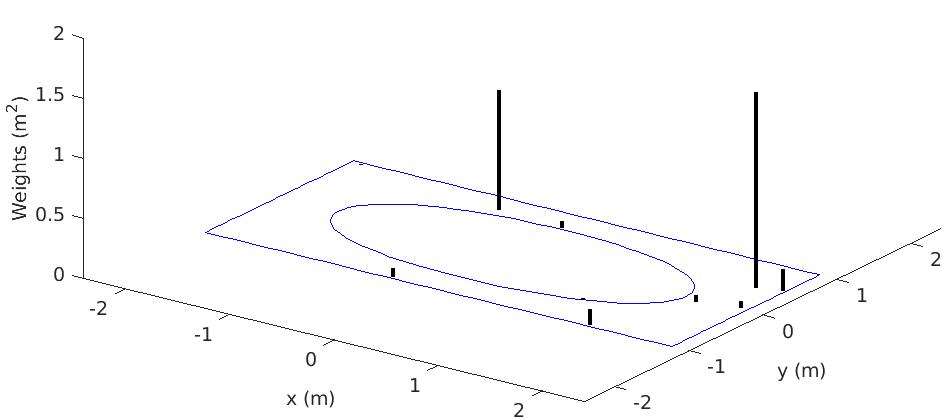}}
  \subfigure[Cluster $1628$  ($\E^C_3 = (0,2,0)$)  ]{\label{fig:cl3}\includegraphics[width=0.49\textwidth]{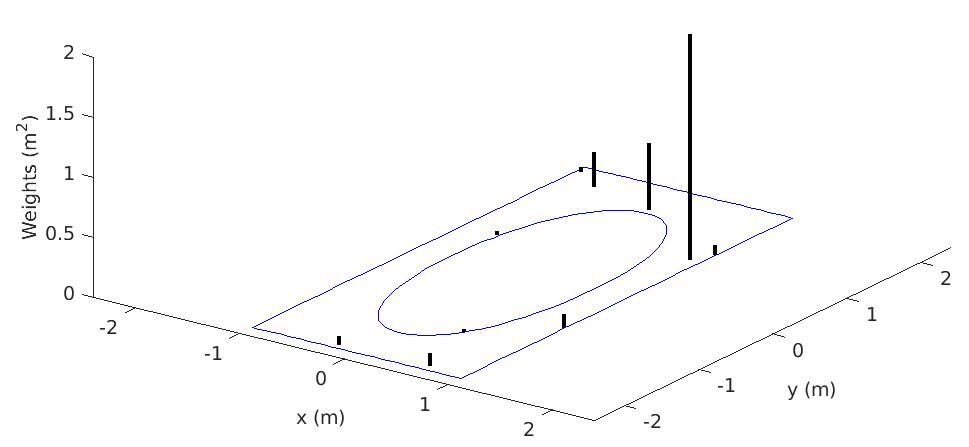}}
  \subfigure[Cluster $2305$  ($\E^C_4 = (-0.2,0,0)$) ]{\label{fig:cl4}\includegraphics[width=0.49\textwidth]{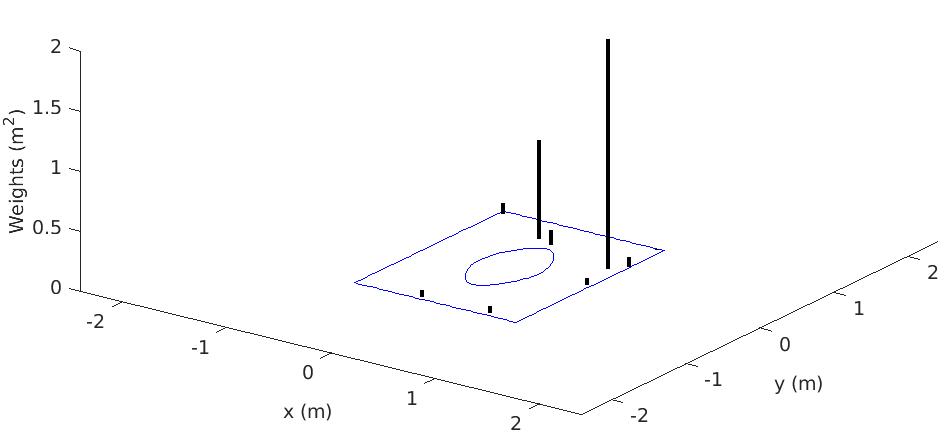}}
  \subfigure[Cluster $2534$  ($\E^C_6 = (0,0,-0.2)$) ]{\label{fig:cl5}\includegraphics[width=0.49\textwidth]{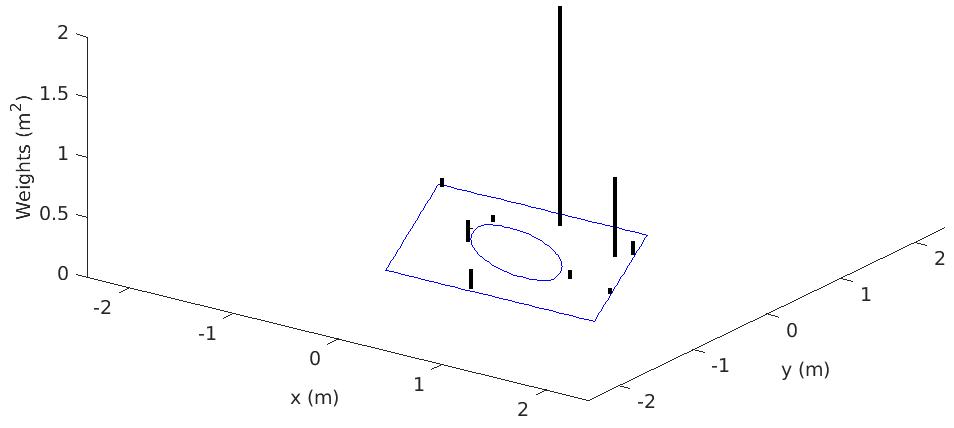}}
  \subfigure[Cluster $3495$ ($\E^C_7 = (2,2,0.2)$)  ]{\label{fig:cl6}\includegraphics[width=0.49\textwidth]{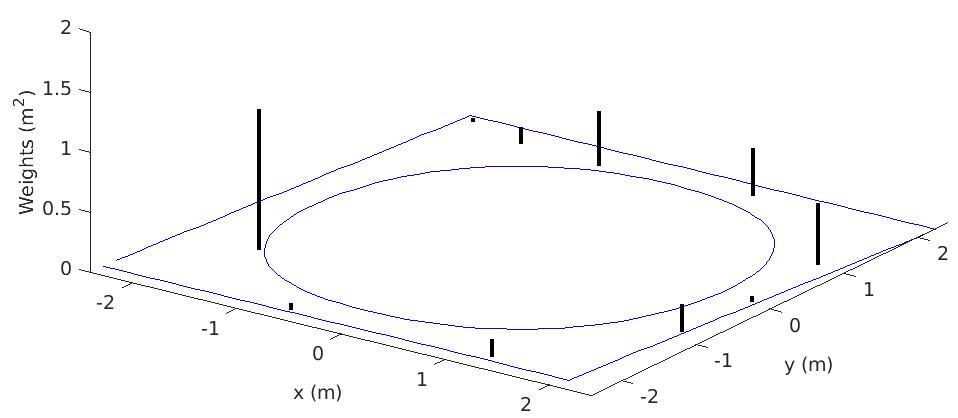}}
   \caption{ Case with $k=3495$ clusters.  Cubature rules (location of points in the deformed configuration  and associated   positive weights, in $m^2$) of the clusters   corresponding to the    input parameters $\E^C_i$ ($i=1,2,3,4,6,7$) (these inputs parameters  define the end points of the piecewise affine training trajectory displayed previously  in Figure \ref{fig:FIG_MESH}.b).    The number of common integration  points for all the clusters is $m=20$; the location of all such common points in the undeformed domain are displayed in Figure \ref{fig:loc}.       }
  \label{fig:FIG_WEIGHTS}
\end{figure}

     Let us now concentrate on the performance on the SAW-ECM of Alg.  \ref{alg:ecm_kratos_ECM_local_pod} when given the integrand basis matrices $\U^1,\U^2 \ldots \U^k$ computed above. As we discussed  in Section \ref{sec: non-uniqueness of ECM},  the solution of  the enhanced ECM in Algorithm \ref{alg:ecm_global} is not unique ---different input candidate points  elicit  different solutions.   As a consequence, the SAW-ECM itself   produces   different outputs  depending on the order in which   $\U^1, \U^2 \ldots  \U^{k}$ are processed.  To examine  the influence of such ordering in the final number of common points, we invoke  the  loop in  Line \ref{alg:2ECM} of  Alg.  \ref{alg:ecm_kratos_ECM_local_pod}   using   100 random permutations of the indices $\{1,2 \ldots 3495\}$.    The resulting discrete probability distribution is displayed in Figure \ref{fig:calls}. Observe that the variations are not quite pronounced (minimum equal to 20, maximum equal to 26), and furthermore, the minimum value is by far  the most frequent output (80 \%).      To gain further insight into how the algorithm selects these  points in one of the permutations leading to $m=20$ points, we plot in Figure \ref{fig:iter} the cardinality of the candidate set with respect to the number of iterations in  the loop of Line  \ref{alg:2ECM} of  Alg.  \ref{alg:ecm_kratos_ECM_local_pod} . It is noteworthy that the final set of points is achieved after only 75  iterations (out of $k=3495$\own{)}.  The location of such  points   is displayed in Figure \ref{fig:loc}: approximately 40\% of the points are relatively near the inner boundary, 40\% relatively close to the outer boundary, and the rest are scattered throughout the domain.

\begin{figure}[!ht]
  \centering
  \subfigure[ ]{\label{fig:p1}\includegraphics[width=0.49\textwidth]{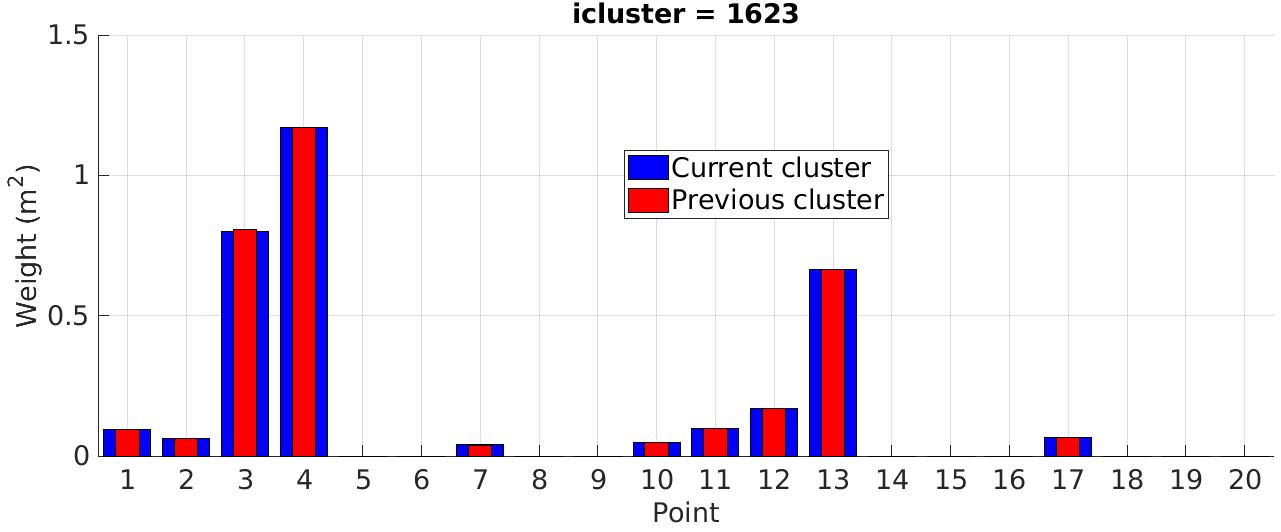}}
         \subfigure[  ]{\label{fig:p2}\includegraphics[width=0.49\textwidth]{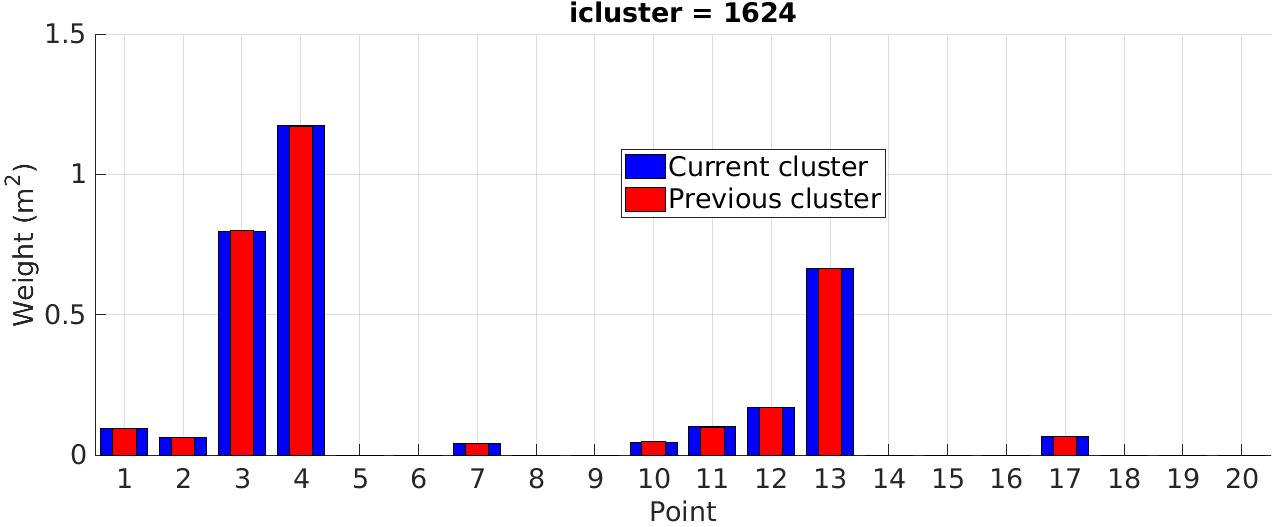}}
  \subfigure[   ]{\label{fig:p3}\includegraphics[width=0.49\textwidth]{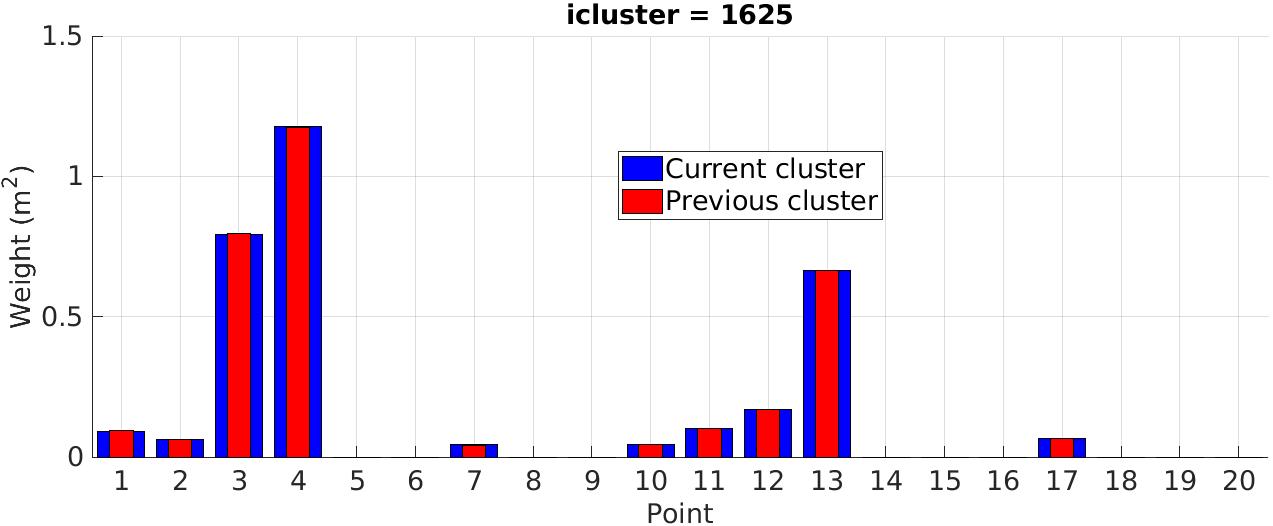}}
  \subfigure[  ]{\label{fig:p4}\includegraphics[width=0.49\textwidth]{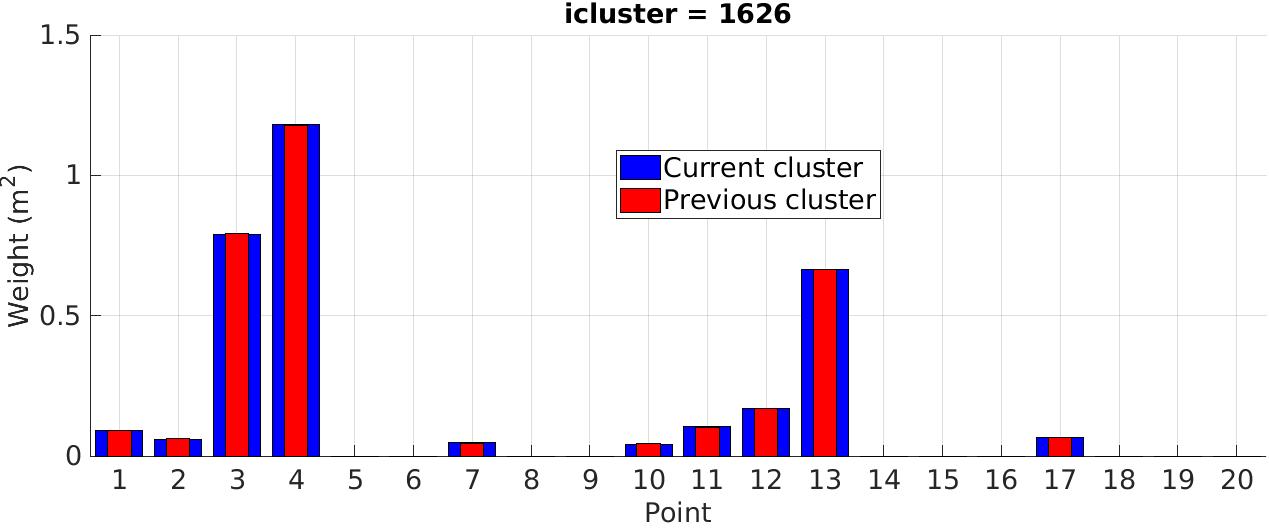}}
  \subfigure[ ]{\label{fig:p5}\includegraphics[width=0.49\textwidth]{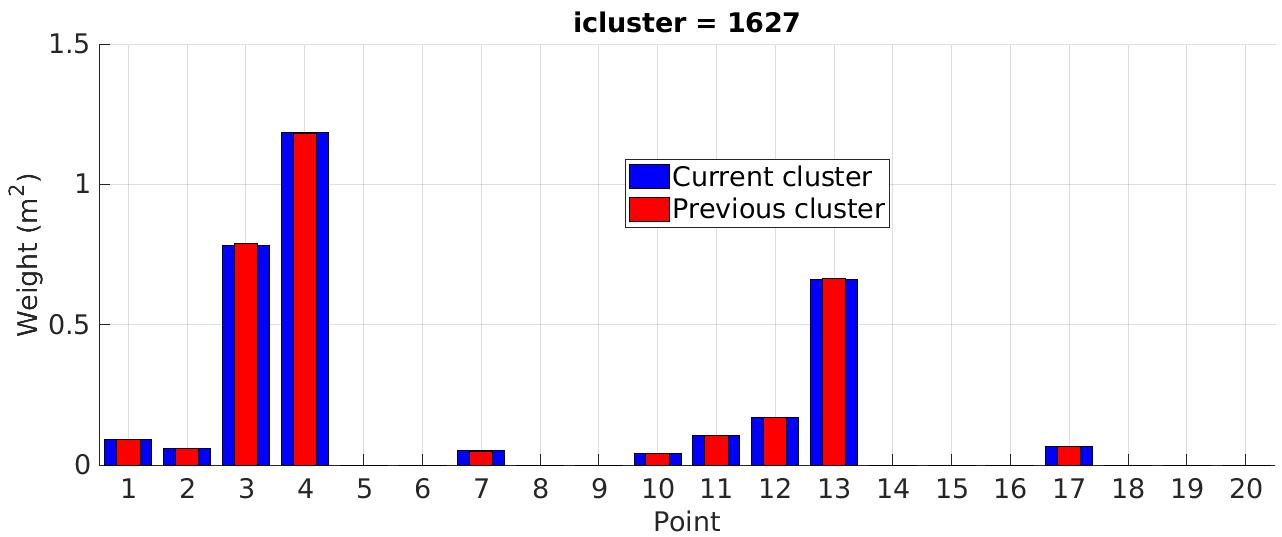}}
  \subfigure[  ]{\label{fig:p6}\includegraphics[width=0.49\textwidth]{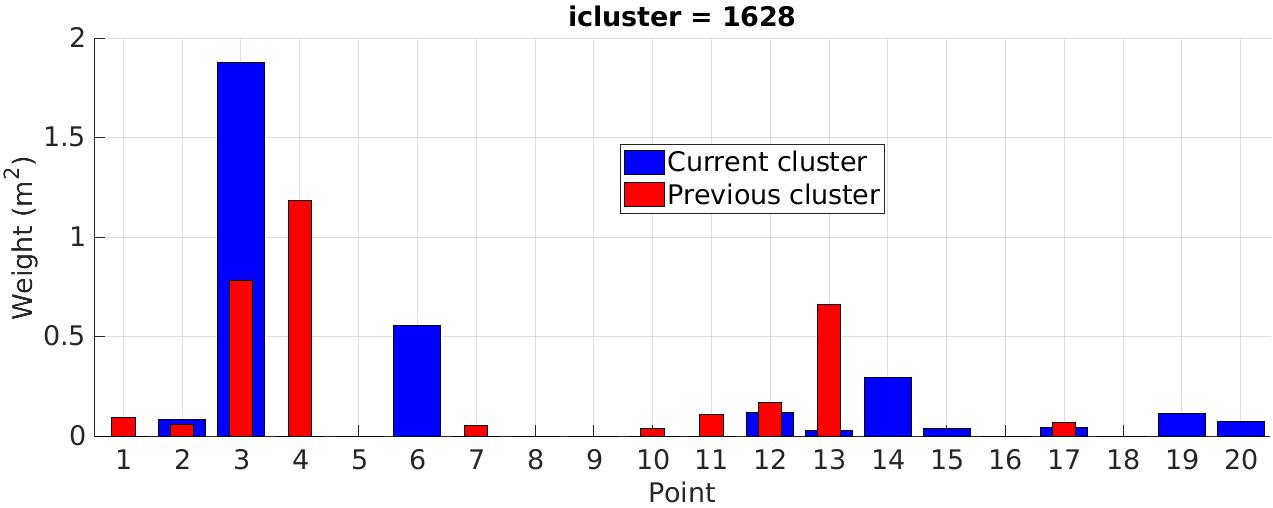}}
   \caption{Case with $k=3495$ clusters.  Cubature rules corresponding  to clusters 1623 to 1628 ---these are the last 6 clusters of the trajectory going from input parameter $\E^C_2 = [2,0,0]$ to $\E^C_3 = [0,2,0]$ in Figure \ref{fig:FIG_MESH}.b. The point indices from 1 to 20 correspond to the labels shown in Figure \ref{fig:loc}.      }
  \label{fig:FIG_TRANSW}
\end{figure}

     Recall that these $m=20$ points   are the points common to all the $k=3495$ subspaces. Each subspace has in principle a different distribution of weights, with a maximum of $m_{max}=10$ nonzero weights for each cluster.  This can be appreciated in  Figures \ref{fig:FIG_WEIGHTS} and   \ref{fig:FIG_TRANSW}. In the former,   we
      display the cubature rules corresponding to the  clusters whose centroids are the end points of the piecewise affine training trajectory defined previously in Figure \ref{fig:FIG_MESH}.b (along with the corresponding deformed shapes).
        In Figure \ref{fig:FIG_TRANSW}, on the other hand, we  show   bar plots of the  weights for a sequence of 6 consecutive clusters (from clusters  1623 to  1628).
       These bar plots reveal that, even though the basis matrices $\U^1, \U^2 \ldots \U^k$ are  processed in random order, and therefore, the SAW-ECM of Alg. \ref{alg:ecm_kratos_ECM_local_pod}  has no information about how close in the solution space \own{such clusters are}, some consecutive subspaces   do display  similar weights distribution.     For instance, from cluster  1623 (Fig. \ref{fig:p1})   to cluster 1627 (Fig. \ref{fig:p5}), the set of 10 indices with nonzero weights is the same, and, furthermore,  variations in weights are  practically imperceptible. This is in contrast to  the transition from cluster 1627 to 1628 (Fig. \ref{fig:p6}), which   is  clearly more abrupt: weights corresponding to  points whose local indices are $\{1,4,7,10,11\}$ become zero, while points $\{6,14,15,19,20 \}$ become active. This might be due to the fact that   the centroid of  cluster 1628    marks the transition from a  tension loading state in the $y$-direction, to a compressive state in the $x$-direction (see  Figures \ref{fig:cl3} and \ref{fig:FIG_MESH}.b).

 Once the basis matrices for displacements\footnote{ {In order to avoid storing the $k=3495$ basis matrices $\PhiB^1,\PhiB^2 \ldots \PhiB^k$, it proves useful to determine a global basis matrix $\PhiB^{all}$ for the column space of the snapshots, and then pre-compute and store the products  $\betaB^i = {\PhiB^{all}}^T \PhiB^i$ ($i=1,2 \ldots k$). This way the nodal displacements may be reconstructed by making $\d^t = \PhiB^{all} \betaB^T \q^t$  }} $\PhiB^1,\PhiB^2$ and $\PhiB^k$, as well as the   location of the  $m=20$ points and the corresponding $k=3495$ sets of $m=20$  weights have been determined in the offline stage,  we can proceed to construct the online local HROMs, and  examine its ability to accurately reproduce the same nodal displacements  used for training the model ---as done in Section \ref{eq:9.222222} for the case of $k=1$ cluster.

  At each time step $t$,     the   local HROM  entails solving   a   system of  nonlinear equations ${\PhiB^{i}}^T \Fint(\q^{i}) = \zero$, which is assembled at each iteration via hyperreduction (see Eq. \ref{eq:9dsd..d*d}).
    Here the index $i$ refers to the cluster which is active at this time step; it can be readily seen that,  in this        limiting case  of $k=3495$ subspaces and $P = 3497$ snapshots,   $i = t$ for $1 < t < P $ (cluster transition takes place at each time step). Likewise, as   pointed out  in Section \ref{sec:trans},  the initial guess for this nonlinear system of equations is obtained from the reduced coordinates of the previous time step: $\q^{t+1}_0 = (\PhiB^{{t+1}^T} \PhiB^t) \q^t$.   Notice  that the
  product ${\PhiB^{t+1}}^T \PhiB^t$ can be pre-computed in the offline stage as well. Thus,  all the online operations of the local HROM are independent of the size of the underlying finite element mesh: the number of nonlinear equations is equal to $n_i= 2$ or $3$ depending on the cluster (see Figure \ref{fig:FIG_MODES}), and the number of points at which the 1st Piola-Kirchhoff  stress and the corresponding constitutive tangent matrix have to be computed at each iteration is equal to $m=20$.

Once we have computed the reduced coordinates $\q^1,\q^2 \ldots \q^P$, in a  post-process stage,  we      reconstruct the nodal displacements as $\d^t = \PhiB^t \q^t$ and compute the Frobenious norm of the difference between the matrices of full-order FE snapshots and HROMs snapshots (divided by the norm of the former).    This gives an  error of $0.88 \cdot 10^{-8}$, which is  slightly below  the employed truncation threshold for the local SVD, $\epsilon = 10^{-8}$. Therefore, we can conclude that  the local HROM constructed with the proposed 20-points  subspace-adaptive weight cubature rule is, not only consistent, but even more accurate than the global HROM described  in Section \ref{eq:9.222222}, which recall required a cubature rule with  $3443$ integration points. \revtwo{As for  the online computational cost, the local hyperreduced-order model (HROM) requires approximately 20 seconds to complete the 3500 time steps, using an in-house MATLAB code executed on an Intel(R) Core(TM) i7-8700 CPU at 3.20GHz with 64 GB RAM on a Linux platform. In comparison, the full-order model takes about 300 seconds, while the global HROM discussed in Section \ref{eq:9.222222} requires 220 seconds.}

The benefits  of using the local HROM in combination with subspace-adaptive weights  are not evident only in the online stage. The offline stage is considerably less  expensive in terms of both computational time and memory requirements as well.    For instance, in the   the case of the global HROM addressed in Section \ref{eq:9.222222}, the computing time required for the ECM of  Alg. \ref{alg:ecm_global} to select the $3443$ points and  weights    is  330 seconds ( using an in-house Matlab code  executed in an Intel(R) Core(TM) i7-8700 CPU, 3.20GHz with 64 Gb RAM, in a Linux platform).  By contrast, the selection of one single set of $m=20$   points (and their corresponding $k=3495$ weights) via the SAW-ECM of Alg. \ref{alg:ecm_kratos_ECM_local_pod} using the same computer  takes  approximately  20 seconds.

        \begin{figure}[!ht]
    \centering
    \includegraphics[width=0.8\linewidth]{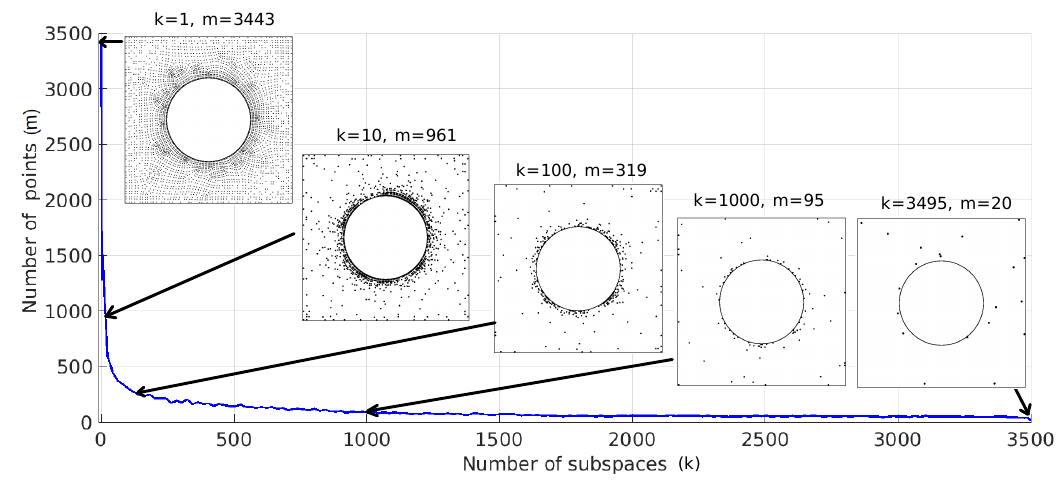}
    \caption{ Number of points determined by the SAW-ECM versus number of subspaces (clusters), along with the distribution  of  integration points for number of clusters equal to $k=1,10,100,1000,3495$.   }
    \label{fig:FIG_evolw}
\end{figure}

\subsubsection{Parameter study in terms of number of subspaces}

In the preceding two sections we have studied the   limiting cases of minimum number ($k=1$) and maximum number ($k=3495$) of subspaces for the given training trajectory.  By way of conclusion, we present in Figure \ref{fig:FIG_evolw}  the number of points obtained by the SAW-ECM (with 10 random permutations per case) for intermediate values of the number of  clusters, along with the distribution of  integration points   for the cases $k=1,10,100,1000,3495$. It can be seen  that the decay is more pronounced in the first portion of the curve ($k=1$ to $k=100$), and then gradually diminishes until achieving the minimum number of points, which in this case is $m=20$. Inspection of the distribution of  points  as the number of subspace increases reveals that the decrease of points is not spatially uniform, but tends to be more pronounced away from the inner boundary, in which, not coincidentally (stress gradients are higher in this area) most of the points concentrate.

\jahoHIDE{It also showcases the potential of the method ....
Physical interpretation ...Compelling interpretation...The training   manifold is P = 3495 ....Is a curve, intrinsic dimensions ...Clustering ....represents as the collection of subspaces ...This means that the integrand function also lives in a subspace of the dimension 1. However, the necessity of overlapping for ensure makes $\gamma = 1, 22$}

  \jahoHIDE{\input{HROM_raul}}

 \section{Concluding remarks}
\label{sec:Conclusions}

 We have formulated and solved the (as yet unaddressed) problem of    deriving quadrature/cubature    rules able to deal   with  multiple subspaces of   functions with integration points common to all the subspaces, yet different set of weights for each subspace.    We have proposed two possible approaches (SAW-ECM and linear-programming based approach), and we have shown that   the former is  less computationally costly and more efficient  (less integration points for the same accuracy) than the latter.   The reason for the higher efficiency of the SAW-ECM is not clear though. One plausible explanation could be that the SAW-ECM   takes advantage of the requirement that the full-order  weights be strictly positive ---the concept of conical hull introduced in Section  \ref{sec: non-uniqueness of ECM} is based on this very hypothesis--- whereas the linear-programming strategy does not use this property in any of its steps.

  Concerning the application of the SAW-ECM in the construction  of   local hyperreduced-order models, we have demonstrated that the use of  adaptive weights does not introduce additional errors, in the sense that if the truncation tolerance of the SVDs for determining basis matrices for the subspaces is set to zero, then the a posteriori error is due exclusively to the first reduction stage (and not to the hyperreduction per se).    We have seen as well that the gains in computational performance increases notably with the  number of subspaces used to represent the region  of the  solution  manifold corresponding to the training parameters.     For instance, in the studied case,  the limiting case of considering each displacement snapshot the centroid of a  cluster  furnishes a hyperreduction factor of over 200   with respect to the standard approach of using the same weights   for all the clusters ($m=20$ points versus $4020$ points ).

  On the other hand,  it may be   inferred from the results obtained in this   limiting case  that the   minimum number that can be achieved with the proposed SAW-ECM is dictated by both  the intrinsic dimensionality of the integrand manifold and the level of overlapping necessary to avoid transition errors between subspaces.   Indeed,   in the studied problem,  we have used   only one piecewise affine trajectory in parameter space;  since  the problem is quasi-static and the constitutive equation is nonlinear elastic (it has no internal variables, and thus, no path-dependence),   the projection of this  trajectory in  the solution manifold is a piecewise continuos curve, and therefore, its intrinsic dimensionality is  $\gamma = 1$. We have argued that the minimum degree of overlapping required for avoiding transition errors  in reproducing this very same trajectory  is $\alpha_{over} = 2$, and that  the integrand  (internal work per unit volume) scales locally with the square of the displacements. Under such circumstances, it follows that the maximum dimension of the integrand subspaces is  $m_{max} =  (\gamma + \alpha_{over})^2 +1 = 10 $ (the additional one accounts for the integration of the volume), and therefore,  the number of points determined by the SAW-ECM when hyperreducing   internal forces  is given by $m = k m_{max} = k ( (\gamma + \alpha_{over})^2 +1)$, where $k\ge1$ measures the effficency of the selection process (we have seen that this constant $k$ depends on the order in which the basis matrices are processed; for the studied problem  $k \sim 2$).

  An  objectionable aspect, when viewed from the machine learning perspective,  of the assessment  presented in Section \ref{sec:localH} is   that we have only studied the consistency of the local HROM, that is, its capability to reproduce the same results used for training,  but not   its capability to generalize (reproduce results not observed in training).    As   explained in Section \ref{sec:trans}, this has been  done in order to   analyze solely the component of the error due to the approximation of the spatial integral, and in doing so, eliminate the part which may be attributed to the so-called transition errors, which affect the accuracy of  local HROMs to a lesser or greater extent in a general case (depending on the degree of overlapping between clusters).     Moreover,  the success of  overlapping criteria, no matter how sophisticated, is contingent in turn on how representative of the solution manifold is the ensemble of snapshots collected during the training stage ---another general challenging issue when constructing reduced-order models.  In summary,  the subspace-adaptive weights cubature method proposed here covers only one facet of local hyperreduction (that of selecting the  the minimum number of points for guaranteeing a user-prescribed accuracy level, provided that other error sources are kept in check). To gain  further confidence in the predictive capabilities of the local HROMs based on subspace-adaptive weights,     further research should be devoted to study the interaction between these various sources of errors, in particular, how     the overall transition errors are affected by the number and  location  of the integration points determined by the SAW-ECM.

\section*{Acknowledgements}

The authors   acknowledge the support of  the European High-Performance Computing Joint Undertaking (JU) under grant agreement No. 955558 (the JU receives, in turn,  support from the European Union’s Horizon 2020 research and innovation programme and Spain, Germany,  France, Italy, Poland, Switzerland, Norway), as well as   the R\&D project PCI2021-121944, financed by MCIN/AEI/10.13039/501100011033 and by the ``European Union NextGenerationEU/PRTR''.
J.A. Hern\'{a}ndez  expresseses gratitude by the  support of, on the one hand,   the   ``MCIN/AEI/10.13039/501100011033/y \emph{por FEDER una manera de hacer Europa}'' (PID2021-122518OB-I00), and, on the other hand, the European Union's Horizon 2020 research and innovation programme under Grant Agreement No. 952966 (project FIBREGY).  Lastly, both J.R. Bravo and S. Ares de Parga acknowledge the \emph{Departament de Recerca i Universitats de la Generalitat de Catalunya} for the financial
support through doctoral grants FI-SDUR 2020 and  FI SDUR-2021, respectively.

\bibliographystyle{abbrv}
\bibliography{main.bib}

\appendix

\jahoHIDE{\input{KMEANS}}

\section{Properties of the ECM }
\label{sec:ecmp}

\subsection{Incorporation of a \own{n}on-trivial \own{c}onstraint}
\label{sec:appendix 2}

When the trivial solution, $\boldsymbol{\omega} = \boldsymbol{0}$, is feasible, it is crucial to introduce an additional constraint. This can be readily accomplished by ensuring that the space of  constant functions is contained in the column space of the basis matrix $\boldsymbol{U} \in \mathbb{R}^{M \times m}$.

Given $\boldsymbol{c}:=\{c\}^{\own{M}}$, where $c \in \own{\mathbb{R} \setminus \{0\}}$, the process requires appending the projection of this constant vector onto the column space of $\boldsymbol{U}$:
\begin{equation}
    \boldsymbol{U} \leftarrow \begin{bmatrix}
        \boldsymbol{U} & \boldsymbol{\lambda}^T
    \end{bmatrix},
    \hspace{10mm}
    \boldsymbol{\lambda} := \frac{(\boldsymbol{I} - \boldsymbol{U}\boldsymbol{U}^T)\boldsymbol{c}}{\norm{(\boldsymbol{I} - \boldsymbol{U}\boldsymbol{U}^T)\boldsymbol{c}}} \ .
\end{equation}

A detailed discussion on the rationale behind this modification can be found in  Ref.  \cite{hernandez2024cecm}, section 2.2.

\subsection{Relevance of the $L_2(\Omega)$ \own{o}rthogonality of \own{c}ontinuous \own{ba}sis \own{f}unctions}
\label{sec:appendix 3}

In our preceding studies \cite{hernandez2017dimensional, hernandez2020multiscale}, we consistently employed orthogonal basis functions at the continuum level. In this paper, we have chosen to relax this orthogonality constraint to facilitate a clearer exposition. Furthermore, when our focus shifts to the selection of elements over Gauss-points,  the previous approach is not straightforwardly applicable.

The problem addressed by the ECM is formulated not in terms of the function $\boldsymbol{a}:(\boldsymbol{x},\boldsymbol{\mu}) \mapsto (a_1, \ldots, a_n)$, but relative to a $\boldsymbol{\mu}-$independent basis function, which is discretely known at the Gauss points:

\begin{equation}
    \begin{aligned}
       \boldsymbol{u}: \ & \Omega \rightarrow \mathbb{R}^{\ell} \\
       : \ & \boldsymbol{x} \mapsto (u_1, u_2, \dots, u_\ell)\ .\\
    \end{aligned}
     \label{eq:basis functions}
\end{equation}

Given the assumption of $L_2(\Omega)$ orthogonality on these functions, we can express:

\begin{equation}
    \int_{\Omega} u_i u_j \ d \Omega = \delta_{i j} \quad \text{for} \ i,j = 1,2, \dots, \ell \ .
\end{equation}

This assumption leads to the Gauss integration rule:
\begin{equation}
      \int_{\Omega} u_i u_j \ d \Omega = \sum_{e=1}^{N_{el}} \int_{\Omega^e} u_i ( \boldsymbol{x} ) u_j ( \boldsymbol{x} )  d \Omega^e = \sum_{e=1}^{N_{el}} \sum_{g=1}^{r} u_i ( \boldsymbol{x} ) u_j ( \boldsymbol{x} )  W_g^e \ ;
    \label{eq: gauss quadrature u u }
\end{equation}
which can be compactly written as:
\begin{equation}
    \tilde{\boldsymbol{U}}^T \texttt{diag}(\boldsymbol{W}) \tilde{\boldsymbol{U}}  = \boldsymbol{I} \ ,
\end{equation}
Here, $\tilde{\boldsymbol{U}}$ is a matrix containing the discrete representation at the Gauss points of basis functions which in the continuum comply with $L_2(\Omega)$ orthogonality. Moreover, the operator \texttt{diag}$(\cdot)$ constructs a matrix with its argument's elements on the diagonal and zeroes elsewhere. Thus, asserting orthogonality on the basis functions leads to an orthogonality condition on the matrix representing the basis function discretely at the Gauss points, not in the Euclidean sense, but according to a $\texttt{diag}(\boldsymbol{W})-$norm.

Since the ECM algorithm necessitates input in the form of an orthogonal matrix within the standard Euclidean inner product, a method initially adopted in our prior works involved the following steps to generate an orthogonal matrix that also considers the basis functions' inner product (this is called the weighted SVD):

\begin{mybox2}
 \label{box: A to U}
    \begin{enumerate}
        \item \textbf{Weighted matrix construction} \\
        From the function samples matrix  $\boldsymbol{A}$, compute a weighted matrix $\bar{\boldsymbol{A}}$ as
            \begin{equation}
             \bar{ \boldsymbol{A} } : =   \texttt{diag}\left( \sqrt{\boldsymbol{W}} \right)  \boldsymbol{A}
             \label{eq: weighted matrix}
            \end{equation}

        \item \textbf{Truncated SVD} \\
        Compute the truncated SVD of the weighted matrix as

        \begin{equation}
            (\bar{\boldsymbol{U}}, \boldsymbol{\Sigma}, \boldsymbol{V})\leftarrow \texttt{SVD}(\bar{ \boldsymbol{A} } , \epsilon_{ \text{\tiny SVD }})
            \label{eq: weighted basis matrix}
        \end{equation}

        \item   \textbf{Matrix preparation for ECM} \\
        Matrix $\bar{\boldsymbol{U}}$ can safely be passed to the ECM algorithm as presented in \cite{hernandez2020multiscale}, since it complies with the standard Euclidean inner product, and moreover it complies that

        \begin{equation}
            \bar{\boldsymbol{U}}^T  \texttt{diag}\left( \sqrt{\boldsymbol{W}} \right) \bar{\boldsymbol{U}}  = \boldsymbol{I} = \tilde{\boldsymbol{U}}^T \boldsymbol{W}  \tilde{\boldsymbol{U}}
        \end{equation}

        where $ \tilde{\boldsymbol{U}} := \texttt{diag}\left( \sqrt{\boldsymbol{W}} \right)^{-1} \bar{\boldsymbol{U}}$.

    \end{enumerate}
\end{mybox2}

For meshes with similarly sized elements, the decision to enforce or skip the $L_2(\Omega)$ orthogonality will yield comparable reduced meshes. However, imposing the $L_2(\Omega)$ constraint on a mesh with varying element sizes will prioritize larger elements.

For users intending to implement the $L_2(\Omega)$ orthogonality, modifications to Algorithm \ref{alg:ecm_global} are in order. Specifically, the objective function needs to be redefined as $\boldsymbol{b} = \boldsymbol{U^T} \sqrt{\boldsymbol{W}}$, and adjustments to the weight computations should be made accordingly. Our repository \url{https://github.com/Rbravo555/localECM} accommodates both strategies.

\end{document}